\documentclass{iopjournal}

\usepackage{graphicx}%
\usepackage{multirow}%
\usepackage{amsmath,amssymb,amsfonts}%
\usepackage{amsthm}%
\usepackage{mathrsfs}%
\usepackage[title]{appendix}%
\usepackage{xcolor}%
\usepackage{textcomp}%
\usepackage{manyfoot}%
\usepackage{booktabs}%
\usepackage{algorithm}%
\usepackage{algorithmicx}%
\usepackage{algpseudocode}%
\usepackage{listings}%
\usepackage{bm}

\begin{document}

\newcommand{\be}{\begin{equation}}
\newcommand{\ee}[1]{\label{#1}\end{equation}}
\newcommand{\bem}{\begin{eqnarray}}
\newcommand{\eem}[1]{\label{#1}\end{eqnarray}}
\newcommand{\eq}[1]{Eq.~(\ref{#1})}
\newcommand{\Eq}[1]{Equation~(\ref{#1})}
\newcommand{\ua}{\uparrow}
\newcommand{\da}{\downarrow}
\newcommand{\g}{\dagger}
\newcommand{\rc}[1]{\textcolor{red}{#1}}

\articletype{Topical Review} 

\title{Theory of ballistic SNS junctions revisited }

\author{Edouatd B. Sonin\orcid{0000-0002-8802-8459}}

\affil{Racah Institute of Physics, Hebrew University of Jerusalem, Jerusalem, Israel}



\email{sonin@cc.huji.ac.il}


\begin{abstract}
The recently suggested theory of planar ballistic SNS junctions demonstrated that in junctions without normal scattering (the case  of equal Fermi velocities and effective masses in all layers) phase gradients in superconducting leads  cannot be ignored. The revision of the previous theory ignoring them  essentially changes the current-phase relation (CPR) of short junctions at $T=0$. It transforms it from forward-skewed to backward-skewed. 

The CPR at temperatures exceeding the energy spacing between  Andreev levels but less than the critical temperature was also calculated. The current at these temperatures is temperature independent and decreases with growing $L$ as  $1/L^4$. The previous theory predicted the current exponentially decreasing with growing $T$ and $L$. 

The SNS junctions with normal scattering (non-equal Fermi velocities in leads and inside the junction) were also analyzed. 
At strong mismatch of Fermi velocities  the planar  ballistic SNS junction becomes a common Josephson junction, which is a weak link and satisfies the Ambegaokar--Baratoff relation.

Implications of the analysis for non-planar junctions (narrow normal bridge between two bulk superconductors) are also discussed. A qualitative  explanation of  experimental observation of backward-skewed CPR in short InAs nanowire bridge junctions
was suggested.

\end{abstract}

\section{Introduction}

Originally the Josephson effect was considered as a result of penetration of the superconducting order parameter into an insulator in a Superconductor--Insulator--Superconductor (SIS) junction or into a normal metal in a Superconductor--Normal metal--Superconductor (SNS) junction (proximity effect). The Josephson effect was considered as a result of overlapping  of superconducting order parameters penetrated from two superconducting leads. This is possible if length of non-superconducting segment of the junction  is not too long  compared with the coherence length. However,   more than a half-century ago   it was demonstrated \cite{Kulik,Ishii,Bard} that if the normal metal segment  is  ballistic  the Josephson coupling is possible without proximity effect  even for rather long SNS junctions.

The pioneer  papers \cite{Kulik,Ishii,Bard}  and many subsequent ones (see most recent Refs.~\cite{JJbk,ThunKink,Green}) used the self-consistent field method \cite{deGen}.  In this method  an effective pairing potential is introduced, which transforms the second-quantization Hamiltonian with the electron interaction into an effective Hamiltonian quadratic in creation and annihilation electron operators. The effective Hamiltonian   can be diagonalized by the Bogolyubov\,\textendash\,Valatin transformation. The effective Hamiltonian is not gauge invariant, and the theory using this Hamiltonian  violates the charge conservation  law. The charge conservation law is restored  if one solves the Bogolyubov\,\textendash\,de Gennes equations  together with the self-consistency  equation for the pairing potential. 

An analytical solution of the Bogolyubov\,\textendash\,de Gennes equations  together with the self-consistency  equation for the pairing potential is hardly possible (excepting, probably, some specially cases).  
Starting from the famous paper of Andreev \cite{And65}, the theory  introduced some concepts and used some assumptions simplifying the analysis:

\begin{figure}[!t]
\centering
\includegraphics[width=.3\textwidth]{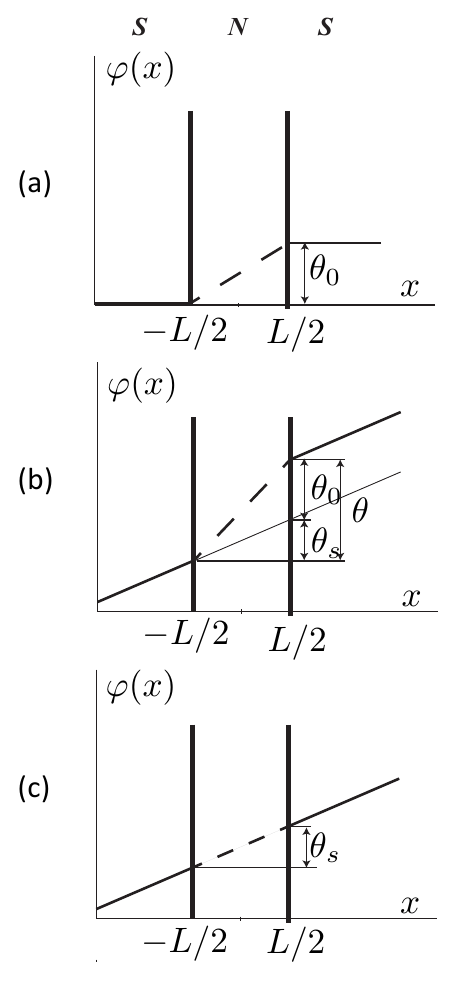}
\caption[]{The phase variation across the SNS junction. (a) The vacuum current produced by the vacuum phase $\theta_0$. The current is confined to the normal layer.   (b) The  superposition of the  vacuum current   and the condensate current determined by the superfluid phase $\theta_s =L\nabla \varphi$. The phase $\theta=\theta_0+\theta_s$ is the Josephson phase. (c) The condensate current produced by the phase gradient  $\nabla \varphi$ in the superconducting layers.  In all layers the electric current is equal to $env_s$.  \label{f1} }
\end{figure}

\begin{enumerate}
\item Instead of solving the self-consistency  equation,  it was assumed   that there is  a gap $\Delta$ of constant modulus $\Delta_0=|\Delta |$  in the superconducting leads and zero gap inside the normal layer. Further we call it the steplike pairing potential model, or, simpler, .the steplike model.

\item  The effective masses and Fermi velocities were assumed to be the same in  all layers. At this assumption, the normal scattering is absent, and  only the Andreev scattering is possible   at SN interfaces.  

\item   Another feature of the case of equal masses and Fermi velocities  was that the original second-order differential Bogolyubov\,\textendash\,de Gennes equations are reduced to the first order equations with boundary conditions imposed only on the wave function, but not on its derivative.

\item  It was assumed that not only the absolute value $\Delta_0$ but also phases were constant in superconducting leads (no phase gradients). Constant phases in two leads, however, were different. 

\end{enumerate}

\begin{figure}[!t]
\centering
\includegraphics[width=.7\textwidth]{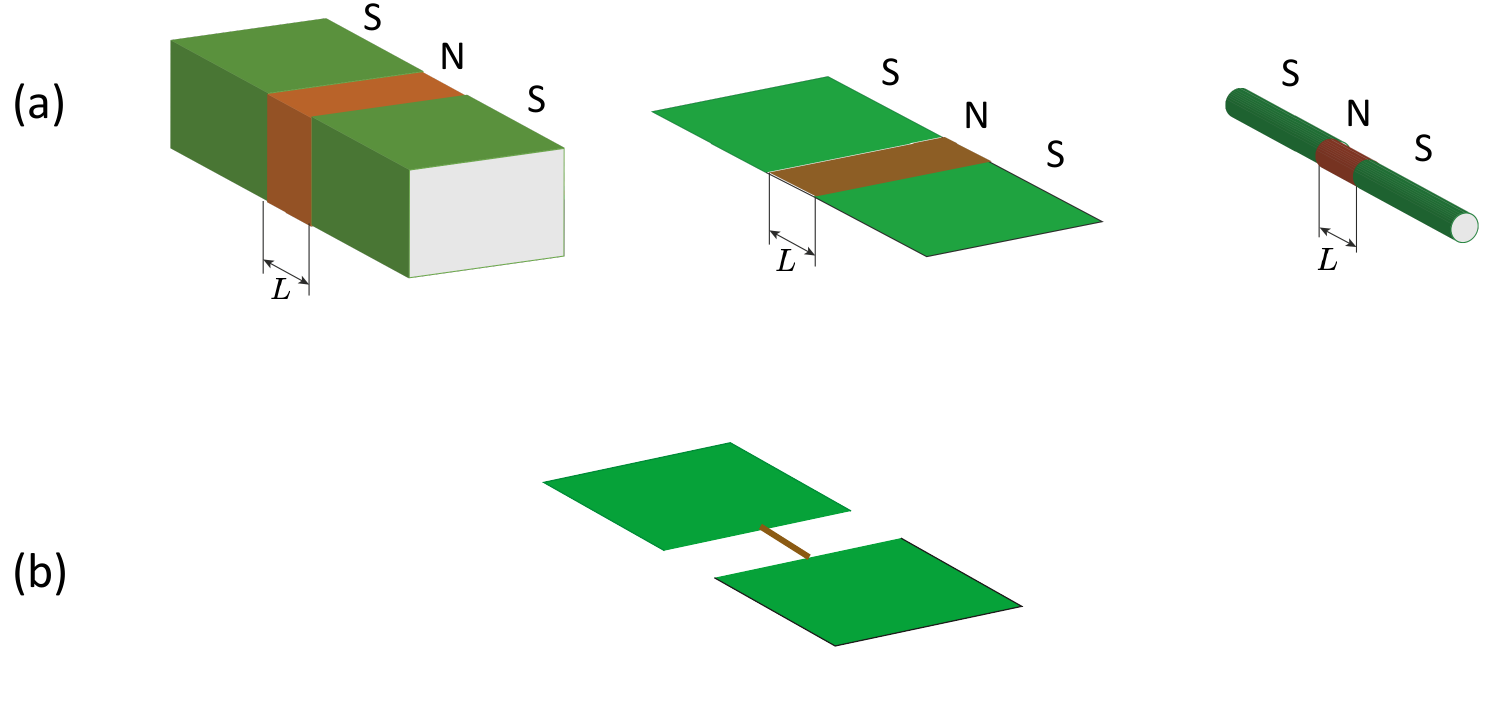}
\caption[]{Geometry of SNS junctions. (a) 3D, 2D, and 1D planar junctions. (b) An example of a non-planar junction: a narrow normal bridge between superconducting leads of higher dimensionality. \label{pg} }
\end{figure}

The last assumption of constant phases in leads is illustrated in Fig.~\ref{f1}(a). At this phase profile, the charge conservation law is violated since the current flows only inside the normal layer. But it was believed that the charge conservation law can be restored by so small phase gradients in leads that this cannot affect calculations  ignoring the gradients. This suggestion is  true for a weak link, inside which the phase varies much faster than in the leads. The assumption was challenged recently \cite{Son21,SonAndr,SonSh,Son26} for planar SNS junctions at $T=0$. In 3D, 2D, and 1D planar junctions  transverse cross-sections of all layers are the same, as shown in Fig.~\ref{pg}(a).  
We shall use the name ``planar junction'' even for junctions in 1D wires when the cross-section  plane becomes a point. In a planar junction at $T=0$ the phase gradients in the leads do affect the current in the normal layer. Thus, one should determine   currents in the normal and  superconducting layers  self-consistently.  

In Refs.~\cite{Son21,SonAndr,SonSh,Son26} and in the present paper the following  gap profile with the constant gradient $\nabla \varphi$ in leads was considered [Fig.~\ref{f1}(b)]:
\be
\Delta =\left\{ \begin{array}{cc} \Delta_0e^{i\theta_0/2+ i\nabla \varphi x} & x>L/2 \\  0 & -L/2<x<L/2\\ \Delta_0e^{-i\theta_0/2+i\nabla \varphi x} & x<-L/2  \end{array} \right. .
   \ee{prof}
The phase profile is determined by two phases $\theta_0$ and $\theta_s=L\nabla\varphi$. The total phase difference across the normal layer (Josephson phase) is a sum of two phases: $\theta=\theta_s+\theta_0$.

In the steplike model there is a mathematically correct analytical  solution of the Bogolyubov\,\textendash\,de Gennes equations for any choice of $\theta_0$ and $\theta_s$. But we filter these solutions by the requirement of the charge conservation law. The strict conservation law was replaced by  a softer condition that, at least,  total currents deep in all layers are the same.

Within the steplike model, the problem is reduced to quadratures. There is the {\em ab initio} exact expression for the current through the junction via the sum over all Andreev bound states and the integral over all continuum states. But for long junctions these sum and integral are rather complicated for analytical and even numerical calculation because of oscillating integrand and necessity to calculate small difference of large terms. The rather sophisticated formalism  of temperature Green function was used \cite{Kulik,Ishii,ThunKink}.

 Among possible solutions of  the  Bogolyubov\,\textendash\,de Gennes equations there is one especially simple. This is for the phase profile as in a uniform superconductor: constant phase gradient $\nabla \varphi$ in all layers [Fig.~\ref{f1}(c)]. In this case $\theta_0=0$,
 and in all layer the same current as in a uniform superconductor flows: 
  \be
J_s=env_s=J_0{\theta_s\over \pi}.
    \ee{Js}
Here $n$ is the electron density, $v_s= {\hbar \over 2m}\nabla \varphi$ is the superfluid velocity,
\be
J_0={ev_f\over L}={\pi en\hbar\over 2mL},
     \ee{JL} 
and $v_f$ is the Fermi velocity. The phase $\theta_s $ and the current  $J_s$ were called the superfluid phase and the condensate current respectively \cite{Son21}. 
 
The state with the condensate current can be obtained by the Galilean transformation of the ground state  without currents ($\theta_s=\theta_0=0$). For long ballistic SNS junctions Galilean invariance  despite broken translational invariance has been already noticed by Bardeen and Johnson \cite{Bard} and used in Ref.~\cite{Son21} in the steplike model.  In Ref.~\cite{SonAndr} the Galilean invariance  was demonstrated for an arbitrary gap profile under the condition that the ratio of the gap  $\Delta_0$ to the Fermi energy $\varepsilon_f$ is small (the weak coupling limit) and  the Andreev reflection is the only mechanism of scattering. This means that state with the condensate current exists beyond  the steplike model  for  junctions with any $L$.

The uniform current state with constant phase gradients was confirmed by numerical calculations by Riedel {\em et al.} \cite{Bagwell}. They did not use the  steplike model and  solved the Bogolyubov\,\textendash\,de Gennes equations together with the integral self-consistency equation for the gap. Riedel {\em et al.} obtained that although the pairing potential amplitude  smoothly varied across the interface between  the normal and superconducting layers, the phase gradient remained strictly constant along the whole junction as in Fig.~\ref{f1}(c).  This phase profile contradicted the existing theory, which assumed that the phase varies only in the normal layer [Fig.~\ref{f1}(a)]. Recently new numerical calculations by Krekels {\em et al.} \cite{Krekels} taking into account the self-consistency equation also confirmed importance of phase gradients in leads.

\begin{figure}[!t] 
\centering
\includegraphics[width=0.4 \textwidth]{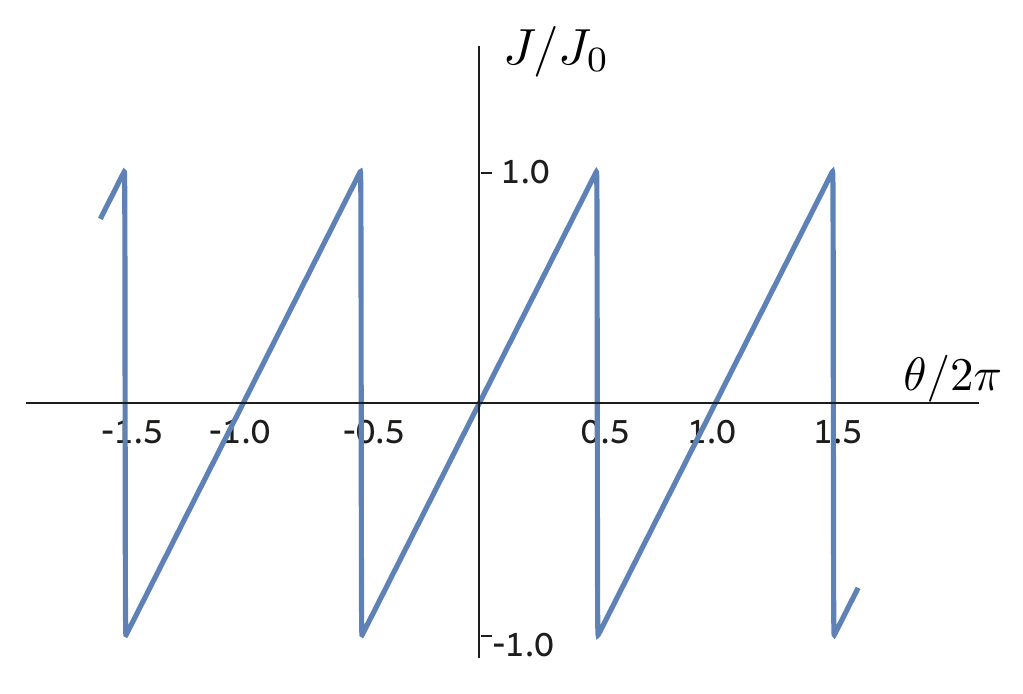} 
 \caption{The saw-tooth CPR at zero temperature. Here $J_0={\pi\hbar\over 2mL}en$ ($={ev_f\over L}$ in the 1D case).\label{ST}}
 \end{figure}

The phase $\theta_0$ and the current $J_v$ produced by this phase were called vacuum phase and vacuum current respectively \cite{Son21}.
The current  $J_v$ flows only in the normal layer at all Andreev  and continuum states  being unoccupied. The charge conservation law is violated, and the current  $J_v$ must be compensated  by the current produced by quasiparticles at Andreev levels, which also    flows only in the normal layer. It was called excitation current $J_q$ \cite{Son21}. At zero temperature the excitation current appears at the critical current determined by the Landau criterion that the energy of the lowest Andreev level reaches 0 \cite{SonSh}. The current-phase relation (CPR) is derived from the condition that the total current flowing only in the normal layer vanishes: $J_v+J_q=0$. Thus, our analysis demonstrated that taking into account phase gradients in the superconducting leads  one can resolve the problem of the charge conservation law {\em within} the steplike model, contrary to what was believed before \cite{ThunKink}.

Despite an essential difference in the physical picture of the charge transport through the junction, for long junctions with $L\gg \zeta_0$ at zero temperature the both theories predicted   the same saw-tooth CPR (Fig.~\ref{ST}), which at $-\pi <\theta<\pi$ is 
\be
J=J_0{\theta \over \pi}.
      \ee{tooth}
Here 
\be
\zeta_0={\hbar v_f\over  \Delta_0} 
     \ee{calL}
 is the coherence length.

Coincidence of the CPRs in the two theories in the limit $L/\zeta_0\to \infty$  led to the wrong conclusion that there is no difference between the condensate and the vacuum currents (see discussion in Refs.~\cite{Thun,Son23rep}). There is a principal difference between two currents and two phases $\theta_s$ and $\theta_0$ producing them. The condensate currents flows in all layers, while the vacuum current flows only in the normal layer violating the charge conservation law. At tuning the phase $\theta_0$ Andreev levels move with respect to the gap sometimes entering into or exiting from the gap, i.e., either a new Andreev level appears, or  an  Andreev level existing before disappears. At tuning the phase $\theta_s$  Andreev levels move together with the gap and their respective positions do not vary.  

In contrast to the limit $L\to \infty$, in the limit $L\to 0$ (short junctions) the difference between the previous theory ignoring phase gradients in leads and the suggested charge-conserving approach  taking them in account becomes very essential \cite{SonSh,Son26}.  Remarkably, at zero temperature it is possible to obtain a simple analytical expression for the CPR for any $L$. In the limit $L\to 0$ the SNS junction becomes a uniform superconductor with a constant current produced by a constant phase gradient without phase jumps. The theory ignoring phase gradients fails to describe this natural  behavior. Neglecting phase gradients, the maximum current  through the junction (critical Josephson current) is always at the phase $\pi$, i.e., the CPR is forward-skewed (the current maximum is located at the phase exceeding the phase $\pi/2$ of the current maximum of the sinusoidal CPR). But taking into account phase gradients, in short  junctions the CPR becomes backward-skewed (the current maximum is located at the phase less than $\pi/2$).\footnote{In the past the backward-skewed CPR was reveled in the theory of dirty SNS junctions at temperatures close to critical  \cite{Ivan} (see also Sec.~V.4 in the review \cite{GKI}). It was explained by depairing (gap suppression determined from the Ginzburg--Landau theory). In contrast, our work  revealed the backward-skewed CPR at $T=0$ in ballistic junctions in the steplike model ignoring the gap suppression in superconducting leads.} 

Krekels {\em et al.} \cite{Krekels} compared their numerical results with our analytical theory and the previous theory. The comparison supports the conclusion that the CPR becomes backward-skewed in short  junction. The backward-skewed CPR was observed in short InAs nanowire junctions by Spanton {\em et al.} \cite{InAs}.  They also acknowledged disagreement with the existing theory,  but  connected this  with Coulomb interaction \cite{InAs2}. Our analysis and numerical calculations \cite{Krekels} show that the backward-skewed CPR is possible also in the model without interaction.

The present article considers also  nonzero temperatures, when one cannot avoid complicated calculations of the sum and the integral determining the vacuum current.  As in the past \cite{Bard,ThunKink}, the $T\neq 0$ analysis focuses on long junctions $L\gg \zeta_0$ when the temperature  exceeds the  Andreev level energy  spacing, but still less than the gap $\Delta_0$ \cite{Son26}.  At short junctions $L \leq \zeta_0$ the temperature effect becomes important at temperatures close to critical. This requires an essential revision of existing theories, which is beyond the scope of the present paper. 

At high temperatures  the total current is very small, and the SNS junction becomes a weak link without essential effect of phase gradients in leads. So, the total current can be approximated by the sum of the vacuum and the excitation current as was done in the past. The  main contribution  $\propto 1/L$ to the excitation current is equal in amplitude but opposite in sign  to the vacuum current in the limit $L\to \infty$. 
In the past the small current was determined by the contribution of Matsubara poles. This contribution takes into account a small difference between the exact sum determining the excitation current in Andrew levels and the main term $\propto 1/L$ obtained by replacing the sum by an integral. The total current  exponentially decreasing with $T$ and $L$.

In Ref.~\cite{SonAndr} it was suggested  that there are other possible  corrections more important at high temperature than the Matsubara contribution. These are power law corrections  $\propto 1/L^w$ to the vacuum current, which  at $w>1$ decrease with growing  $L$ faster than the main   term  $\propto 1/L$. However, corrections $\propto 1/L^{3/2}$ calculated in Ref.~\cite{SonAndr} and corrections $\propto 1/L^2$ calculated in Ref.~\cite{Son26}  vanish in the total current. Failing to find the power law correction analytically, the  current was calculated numerically using the exact {\em ab initio} expressions for currents \cite{Son26} . The calculation revealed a small power-law correction  $\propto 1/L^4$. It is independent from temperature and therefore becomes more important than the exponentially small Matsubara term. As a result,  the critical Josephson current dependence on $T$ has a plateau at temperatures between the   Andreev level energy  spacing and critical one. 

The planar junctions without normal scattering (equal Fermi velocities in all layers) are reviewed in Part \ref{part1} of the article. 
 The present article extended our approach onto the case of non-equal Fermi velocities in superconducting leads and inside the junction (Part \ref{part2}). The CPR in the limit of strong mismatch of Fermi velocities was calculated. It demonstrates  that at strong mismatch of Fermi velocities the planar  ballistic SNS junction becomes a common Josephson junction, which is a weak link and satisfies the Ambegaokar--Baratoff relation connecting the maximal current through the junction with the resistance of the corresponding normal (NNN) ballistic junction. 
 
 In Part \ref{part2} (Sec.~\ref{bnd}) we consider possible implication of our analysis of planar junctions for a narrow normal bridge between superconducting leads of higher dimensionality [Fig.~\ref{pg}(b)] Our quantitative analysis of planar junctions is not directly applicable to this case, but at the qualitative level some conclusions may be done. In particular, it was suggested why Spanton {\em et al.} \cite{InAs} observed a backward-skewed CPR only for a narrow gate voltage range in the gated junction, and never in ungated junctions. .

The analysis in the paper mostly addresses the 1D channel. For the extension of the calculation  on multidimensional systems
currents must be integrated over  transverse components   of   wave vectors.

\part{SNS without normal scattering: equal Fermi velocities in all layers} \label{part1}

\section{Bogolyubov\,\textendash\,de Gennes equations for a moving condensate and Galilean invariance} \label{BdG}

 The  Bogolyubov\,\textendash\,de Gennes equations for the  Bogolyubov\,\textendash\,de Gennes wave function  
\begin{equation}
\psi(x) = \left[ \begin{array}{c} u(x)\\  v(x) \end{array}
\right], 
     \label{spinor} \end{equation}
with the energy $\varepsilon$ are
\bem
\varepsilon  u  =-{\hbar^2 \over 2m} \left( \nabla^2 + k_f^2\right) u  + \Delta  v ,
\nonumber \\
\varepsilon v  ={\hbar^2 \over 2m}  \left( \nabla ^2 + k_f^2\right) v  + \Delta^* u .
     \eem{BG}
Here $k_f$ is the Fermi wave number and $\Delta=\Delta_0(x)e^{i\varphi(x)}$ in general.

 A quasiparticle with wave function \eq{spinor} creates an excitation  current 
 \bem
 j_q=  -{i\hbar\over 2 m}[ (u^*\nabla u-u\nabla u^*) +(v^*\nabla v-v\nabla v^*) ].
 \eem{jq} 
If the state with wave function \eq{spinor} is not occupied, there is a vacuum current
  \bem
j_v = {ie\hbar\over 2m} (v^*\nabla v-v\nabla v^*).
 \eem{jv} 

Since there is no charge conservation law, solution of the  Bogolyubov\,\textendash\,de Gennes equations yields states with charge currents varying  along 
the  SNS junction. But in the Bogolyubov–de Gennes theory there is the conservation law for the total number of quasiparticles, and the quasiparticle flux 
 \be
g= -{i\hbar\over 2 m}[ (u^*\nabla u-u\nabla u^*) - (v^*\nabla v-v\nabla v^*) ]
 \ee{g} 
 must be constant along the planar SNS junction. The conservation law for the total number of quasiparticles follows from the condition that the probability $u^2+v^2$  to find a quasiparticle in some point of  the space after integration over the whole space must be unity.

 If the ratio $\Delta_0/\varepsilon_f$ is very small,  among solutions of the  Bogolyubov\,\textendash\,de Gennes equations
 there are wave functions $\psi(x)$, which  are superpositions of plane  waves with wave numbers only close to either  $+ k_f$, or  $- k_f$. Quasiparticles in these states  will be  called  rightmovers (+) and leftmovers (-) respectively.  After transformation of the wave function,
\be
\left(  \begin{array}{c}
u    \\
  v 
\end{array}  \right) =\left(  \begin{array}{c}
\tilde u    \\
\tilde  v 
\end{array}  \right)e^{\pm  ik_f x},
        \ee{}
the second order  terms   in gradients, $\nabla^2 \tilde  u$ and $\nabla^2 \tilde  v$, can be neglected, and the Bogolyubov\,\textendash\,de Gennes equations are reduced to the equations of the first order  in gradients \cite{And65}:
\bem
\varepsilon \tilde u  =\mp i\hbar v_f \nabla \tilde  u  + \Delta \tilde  v ,
\nonumber \\
\varepsilon \tilde v =\pm  i\hbar v_f \nabla \tilde v  + \Delta^*\tilde  u .
    \eem{BG1}

The boundary conditions at interfaces between layers require the continuity  of the wave function components, but not their gradients.

Let us demonstrate Galilean invariance of the Bogolyubov\,\textendash\,de Gennes equations when the wave functions are superpositions of only rightmovers, or only of leftmovers \cite{SonAndr}. Suppose that we found the Bogolyubov\,\textendash\,de Gennes function    $\left(  \begin{array}{c}
\tilde u_0    \\
\tilde  v_0 
\end{array}  \right)$ with the energy $\varepsilon_0$
 for an arbitrary profile of the superconducting gap $\Delta(x)$. One can check that the  wave function 
 \be
 \left(  \begin{array}{c}
\tilde u    \\
\tilde  v
\end{array}  \right)=\left(  \begin{array}{c}
\tilde u_0 e^{i\nabla\varphi x/2}   \\
\tilde  v_0  e^{-i\nabla\varphi x/2} 
\end{array}  \right)
        \ee{} 
with constant gradient $\nabla\varphi$ satisfies \eq{BG1} where the gap $\Delta(x)$ is replaced by    $\Delta(x)e^{i\nabla\varphi x}$ and the energy  is 
\be
 \varepsilon=\varepsilon_0 \pm   {\hbar^2 k_f\over 2m}\nabla\varphi=\varepsilon_0 \pm   \hbar k_f v_s=\varepsilon_0 \pm    {\hbar^2 k_f\over 2mL}\theta_s .
      \ee{GT}   
Thus, the Galilean transformation produces the same Doppler shift in the energy as in a uniform superconductor. It adds
 the same  condensate current $J_s=env_s$ in all layers, which does not violate the charge conservation law.  Our derivation valid for any profile of the gap in the space and  remains valid if $\Delta$ vanishes in some part of the space.

If there is normal scattering, the wave function is a superposition of rightmovers and  leftmovers. Doppler shifts for them have opposite signs. Therefore, after  the Galilean transformation the wave function is not an eigenstate of the energy. Such wave functions cannot be used in the analysis of the stationary transport.

\subsection{Delocalized continuum scattering  states}

For  a rightmover  quasiparticle and  a leftmover  quasihole  incident from left and  propagating from $x=-\infty$ to $x=\infty$ the wave  function with the energy $\varepsilon$ is
\bem
\left(  \begin{array}{c}
 u    \\
  v 
\end{array}  \right)=\left(  \begin{array}{c}
u_0(\pm \xi) e^{-i\theta_0/4+i \nabla \varphi\,x/2}  \\
 v_0(\pm \xi) e^{i\theta_0/4-i \nabla \varphi\,x/2}
\end{array}  \right)e^{ i\left(\pm k_f+{ m\xi\over \hbar^2k_f}\right)x}
\nonumber \\ 
 +r_\pm \left(  \begin{array}{c}
u_0(\mp \xi)e^{-i\theta_0/4+i \nabla \varphi\,x/2}   \\
 v_0(\mp \xi) e^{i\theta_0/4-i \nabla \varphi\,x/2} 
\end{array}  \right)e^{i\left(\pm k_f-{ m\xi\over \hbar^2k_f}\right)x}
      \eem{BGr}
for $x<-L/2$, 
\be
\left(  \begin{array}{c}
 u    \\
 v 
\end{array}  \right)=t_\pm\left(  \begin{array}{c}
u_0(\pm \xi) e^{i\theta_0/4+i \nabla \varphi\,x/2}  \\
 v_0(\pm \xi) e^{-i\theta_0/4-i \nabla \varphi\,x/2}
\end{array}  \right)
e^{ i\left(\pm k_f+{ m\xi\over \hbar^2k_f}\right)x}
      \ee{BGt}
for $x>L/2$, and 
\begin{equation}
\left(  \begin{array}{c}
 u    \\
 v 
\end{array}  \right)=t_\pm\left(  \begin{array}{c}
u_0(\pm \xi)  e^{i\theta/4 \pm {i m\varepsilon    \over \hbar^2 k_f}x + {i m(\xi \mp \varepsilon ) L\over 2\hbar^2k_f}}   \\  v_0(\pm \xi)   e^{-i\theta/4 \mp {i m\varepsilon    \over \hbar^2 k_f}x +  {i m(\xi \pm \varepsilon ) L\over 2\hbar^2k_f} }
\end{array}  \right) e^{\pm  ik_f x}
  \label{norm}
\end{equation}
 inside the normal layer $-L/2<x<L/2$. Here  $\xi=\sqrt{\varepsilon_0^2-\Delta_0^2}$ and
\be
u_0(\xi) =\sqrt{\varepsilon_0 +\xi\over 2\varepsilon_0},~~v_0(\xi) =\sqrt{\varepsilon_0 -\xi\over 2\varepsilon_0}=u_0(-\xi).
   \ee{u0v0} 
The energy $\varepsilon_0>\Delta_0$ is the energy at resting condensate ($\nabla \varphi =0$), which is connected with the energy  $\varepsilon$ at moving condensate ($\nabla \varphi \neq 0$) by the relation \eq{GT} 
following from the Galilean invariance \cite{SonAndr}. Amplitudes $t_\pm$ and $r_\pm$ of transmission and reflection are determined from the continuity of spinor components at $x=\pm L/2$: 
\bem
t_\pm=\frac{e^{- {i m\xi  L  \over \hbar^2 k_f} }}{\cos\left({\varepsilon m  L\over \hbar^2 k_f}\mp {\theta \over 2}\right)-i{\varepsilon_0\over \xi}\sin\left({\varepsilon m  L\over \hbar^2 k_f}\mp {\theta \over 2}\right)} 
=\frac{e^{- {i m\xi  L  \over \hbar^2 k_f} }}{\cos\left({\varepsilon_0m  L\over \hbar^2 k_f}\mp {\theta_0\over 2}\right)-i{\varepsilon_0\over \xi}\sin\left({\varepsilon_0m  L\over \hbar^2 k_f}\mp {\theta_0\over 2}\right)}   ,
     \eem{r}
\bem
r_\pm ={\Delta_0\over \xi_0}\frac{e^{- {i m\xi  L  \over \hbar^2 k_f} }i\sin\left({\varepsilon m  L\over \hbar^2 k_f}\mp {\theta \over 2}\right)}{\cos\left({\varepsilon m  L\over \hbar^2 k_f}\mp {\theta \over 2}\right)-i{\varepsilon_0\over \xi}\sin\left({\varepsilon m  L\over \hbar^2 k_f}\mp {\theta \over 2}\right)}
={\Delta_0\over \xi_0}\frac{e^{- {i m\xi  L  \over \hbar^2 k_f} }i\sin\left({\varepsilon_0m  L\over \hbar^2 k_f}\mp{\theta_0\over 2}\right)}{\cos\left({\varepsilon_0m  L\over \hbar^2 k_f}\mp {\theta_0\over 2}\right)-i{\varepsilon_0\over \xi}\sin\left({\varepsilon_0m  L\over \hbar^2 k_f}-{\theta_0\over 2}\right)}.
\eem{tt} 

The reflection and the transmission probabilities  are
\bem
{\cal R}_\pm=\lvert r_\pm\rvert^2= \frac{\Delta_0^2\left[1- \cos\left({2\varepsilon m  L\over \hbar^2 k_f}\mp \theta_0\right)\right]} {2\varepsilon_0^2 -\Delta_0^2-\Delta_0^2\cos\left({2\varepsilon_0 m  L\over \hbar^2 k_f}\mp \theta_0\right)},
~~{\cal T}_\pm=\lvert t_\pm\rvert^2= \frac{2(\varepsilon_0 ^2-\Delta_0^2)} {2\varepsilon_0^2 -\Delta_0^2-\Delta_0^2\cos\left({2\varepsilon_0 m  L\over \hbar^2 k_f}\mp \theta_0\right)}.
     \eem{T} 
These expressions are also valid for a leftmover quasiparticle and a rightmover quasihole incident from the right. Scattering probabilities depend only on whether quasiparticle or  quasihole is a rightmover or a leftmover. A transition from rightmover (leftmover)  
 quasiparticle to rightmover (leftmover)   quasihole does not change scattering probabilities, as expected from the particle-hole symmetry. Motion of the condensate has no effect on scattering parameters.
The scattering parameters for continuum states were determined by  Bardeen and Johnson \cite{Bard} without phase gradients and  in Refs.~\cite{Bagw,Son21,EBS} with phase gradients.

The scattering probabilities satisfy the equality ${\cal R}_\pm+{\cal T}_\pm=1$ following from  the  conservation law for the number of quasiparticles.

\subsection{Andreev bound states}

The wave function for Andreev bound states at the energy $0<\varepsilon_0<\Delta_0$ is:
 \begin{equation}
\left(  \begin{array}{c}
 u    \\
  v 
\end{array}  \right)=\sqrt{N\over 2}\left(  \begin{array}{c}
 e^{\pm {i   \eta\over  2} +i \nabla \varphi\,x/2 +i\theta_0/4}    \\
  e^{\mp{i   \eta\over  2}-i \nabla \varphi\,x/2 -i\theta_0/4 } 
\end{array}  \right) e^{\pm  ik_f x-(x-L/2) /\zeta}
          \label{sup2}
\end{equation}
 inside the superconducting layer at $x>L/2$,  
   \begin{equation}
\left(  \begin{array}{c}
u    \\
 v 
\end{array}  \right)=\sqrt{N\over 2}\left(  \begin{array}{c}
e^{\pm {i   \eta\over  2}+i \nabla \varphi\,x/2  -i\theta_0/4 } \\
 e^{ \pm {3i   \eta\over  2} -i \nabla \varphi\,x/2 +i\theta_0/4} 
\end{array}  \right)
e^{\pm  ik_f x +i\theta/2\mp{i m\varepsilon  L  \over \hbar^2 k_f} +   (x+L/2)/\zeta },
\label{sup1}
\end{equation}
inside the superconducting layer at $x<-L/2$, and
 \begin{equation}
\left(  \begin{array}{c}
 u    \\
 v 
\end{array}  \right)
=\sqrt{N\over 2}\left(  \begin{array}{c}
  e^{ \pm{i   \eta\over  2}+i\theta/4 \pm  {i m\varepsilon    \over \hbar^2 k_f}(x-L/2) }   \\    e^{\mp  {i   \eta\over  2} -i\theta/4 \mp {i m\varepsilon    \over \hbar^2 k_f}(x-L/2) }
\end{array}  \right)e^{\pm  ik_f x}. 
  \label{normA}
\end{equation}
 inside the normal layer $-L/2<x<L/2$. Here
\begin{equation}
\eta=\arccos{\varepsilon_0\over \Delta_0}.
 \label{eta} \end{equation}
The normalization constant
\be
N={1\over L+\zeta }
    \ee{}
takes into account  that the bound states penetrate into the superconducting layers on 
the penetration length
   \be
\zeta = \zeta_0{\Delta_0\over \sqrt{\Delta_0^2-\varepsilon_0^2}}.
  \ee{zeta}
The length diverges when  $\varepsilon_0$ approaches to the gap $\Delta_0$.

The wave function for Andreev states in Eqs.~(\ref{sup2})--(\ref{normA}) satisfy the boundary conditions (continuity of wave function components) at $x=L/2$. 
The boundary conditions at $x=-L/2$  are satisfied at the Bohr--Sommerfeld condition, which determines the energy spectrum of the Andreev states:
\be
\varepsilon_\pm (s)={\hbar v_f\over 2L}\left[2\pi  s+2\arccos {\varepsilon_{0\pm} (s)\over \Delta_0} \pm \theta\right].
   \ee{eps0}
or 
\be
\varepsilon_{0\pm} (s)={\hbar v_f\over 2L}\left[2\pi  s+2 \arccos {\varepsilon_{0\pm} (s)\over \Delta_0} \pm \theta_0\right].
   \ee{eps00}
Here $s$ is an  integer varying from zero to maximal value  satisfying the condition that $\varepsilon_{0\pm}<\Delta_0$.

At small energy $\varepsilon_{0\pm} (s) \ll \Delta_0$ (small $s$) 
\be
\varepsilon_{0\pm} (s)={\hbar v_f\over 2L+\zeta_0}\left[2\pi  \left(s+{1\over 2}\right)\pm \theta_0\right],
   \ee{en0}
  or
  \be
\varepsilon_{\pm} (s)={\hbar v_f\over 2L+\zeta_0}\left[2\pi  \left(s+{1\over 2}\right)\pm \theta\right].
   \ee{enG}

For  Andreev levels close to the gap one can  expand the arcsin function in \eq{eps00} in $\Delta_0- \varepsilon_{0\pm}$ transforming it to 
\be
 \Delta_0- \varepsilon_{0\pm}   -{\zeta_0\over L} \sqrt{2\Delta_0(\Delta_0- \varepsilon_{0\pm} )}
=\pi ( t+\alpha){\zeta_0 \over L}\Delta_0.
 \ee{eqD}

Here   $\alpha$ ($0 <\alpha <1 $)  is the parameter of incommensurability, which is the fractional part of the ratio of the gap $\Delta_0$ to the Andreev level energy spacing,
\be
{\Delta_0  L\over \pi \hbar v_f}={L\over \pi\zeta_0}=s_m +\alpha,
     \ee{inc}
 $s_m$ is the maximal integer $s$ less than the ratio, and $t=s_m-s$. Solution of \eq{eqD} quadratic with respect to $\sqrt{\Delta_0- \varepsilon_{0\pm}}$ yields
\bem
\varepsilon_{0\pm} =\Delta_0-{\pi \zeta_0\over L}\Delta_0\left[\sqrt{t +\alpha  \mp  {\theta_0\over  2\pi }+{\zeta_0\over 2\pi L }}-\sqrt{{\zeta_0\over 2\pi L}}\right]^2,
\nonumber \\
\sqrt{\Delta_0^2- \varepsilon_{0\pm} ^2}=\Delta_0\sqrt{2\pi \zeta_0 \over L}\left[\sqrt{t +\alpha  \mp  {\theta_0\over  2\pi }+{\zeta_0\over 2\pi L }}
-\sqrt{\zeta_0\over 2\pi L}\right].
    \eem{Edel}

In the limit $L= 0$ the SNS junction becomes a uniform superconductor without normal layer. Still there is the bound state 
with the energy  determined from \eq{en0} at $s=0$:
\be
\varepsilon_{ps}=\Delta_0\cos{\theta_0\over 2}.
  \ee{ASJ} 
This ``Andreev'' bound state  describes a phase slip center in a uniform superconductor with a phase jump at $x=0$. Later on it will be called the  phase slip state. In this  state there are two  evanescent plane waves at $x>0$ and $x<0$ [Eqs.~(\ref{sup2}) and (\ref{sup1}) at $L=0$]. Its energy $\varepsilon_{ps}$ in \eq{ASJ}  directly follows from the continuity of the wave function at $x=0$ ($\theta_s\propto L \to 0$, $\theta=\theta_0$):
\be
\left(  \begin{array}{c}
 e^{\pm {i   \eta\over  2} +i\theta_0 /4}    \\
  e^{\mp{i   \eta\over  2} -i\theta_0 /4 } 
\end{array}  \right) =
\left(  \begin{array}{c}
e^{\mp {i   \eta\over  2}  -i\theta_0/4 } \\
 e^{ \pm {i   \eta\over  2}  +i\theta_0/4} 
\end{array}  \right)
e^{\pm i\eta +\theta_0/2 }.
      \ee{CE}

\section{{\em Ab initio} expressions for currents }
 \label{ASSNS}

In this section we present  {\em  ab initio} expressions for the vacuum and the excitation currents in the normal layer. The expression \eq{Js} for the condensate current directly follows from Galilean invariance and does not  need further discussion.

\subsection{Vacuum current in continuum states}\label{vacC}

The vacuum current in continuum states inside the normal area does not differ from that in the lead, in which the transmitted wave  propagates. Thus, it is determined by the transmission probabilities ${\cal T}_\pm$. One can transform the expression for ${\cal T}_\pm$ in \eq{T} revealing its  dependence on the incommensurability the parameter $\alpha$ introduced in \eq{inc}:
\begin{equation}
{\cal T}_\pm= \frac{2(\varepsilon_0 ^2-\Delta_0^2)} {2\varepsilon_0 ^2 -\Delta_0^2-\Delta_0^2\cos\left[{2(\varepsilon_0 -\Delta_0)m  L\over \hbar^2 k_f}+2\pi \alpha \pm \theta_0\right]}.
     \label{Ta} \end{equation}
At large $L$ the transmission probability  rapidly oscillates  as a functions of  energy.

Collecting together contributions from rightmovers and leftmovers, quasiparticles and quasiholes, the  {\em ab initio} expression for  the continuum vacuum  current is \cite{Son21}
\be
J_{vC} ={e\over \pi\hbar}\int_{\Delta_0}^\infty  ({\cal T}_--{\cal T}_+ ) d\xi.  
     \ee{cJ}

At $L\to 0$ the current $J_{vC}$ vanishes since ${\cal T}_-={\cal T}_+ $ in this limit.

\subsection{Vacuum current in bound Andreev states}

 The current in the Andreev state is determined by the canonical relation connecting it with the derivative of the  energy with respect to the phase:
\be
j_\pm(s)={2e\over \hbar}{\partial  \varepsilon_{0\pm} (s)\over \partial \theta_0}=\pm{e\over \pi \hbar}{\partial  \varepsilon_{0\pm} (s)\over \partial s}=\pm{ev_f\over L+\zeta}
=\pm {ev_f\over L}\frac{\sqrt{1- {\varepsilon_{0\pm} (s)^2\over \Delta_0^2}}}{\sqrt{1- {\varepsilon_{0\pm} (s)^2\over \Delta_0^2}}+{\zeta_0\over L}}.
       \ee{js}
 The factor 2 takes into account that $\theta_0$ is the phase of a Cooper pair but not of a single electron.

The current $j_\pm(s)$ is a current produced by a quasiparticle created at the $s$th state. If the state is not occupied, the vacuum current is two times less than $j_\pm(s)$, and  has an opposite sign. Taking this into account together with two spin states, the {\em ab initio} expression for the vacuum current in  bound states is 
\bem
J_{vA}=-{e\over \pi \hbar}\sum_s \left\{{\partial  \varepsilon_{0+} (s)\over \partial s}\mbox{H}[\varepsilon_{0+} (s)]  \mbox{H}[\Delta_0 -\varepsilon_{0+} (s)] 
\right. \nonumber \\  \left.
-{\partial  \varepsilon_{0-} (s)\over \partial s}\mbox{H}[\varepsilon_{0-} (s)]  \mbox{H}[\Delta_0 -\varepsilon_{0-} (s)]\right\}
\nonumber \\
=-{ev_f\over L}\sum_s  \left\{\frac{ \sqrt{1- {\varepsilon_{0+} (s)^2\over \Delta_0^2}} }{\sqrt{1- {\varepsilon_{0+} (s)^2\over \Delta_0^2}} +{\zeta_0\over L}}\mbox{H}[\varepsilon_{0+} (s)]  \mbox{H}[\Delta_0 -\varepsilon_{0+} (s)]
\right. \nonumber \\  \left.
- \frac{ \sqrt{1- {\varepsilon_{0-} (s)^2\over \Delta_0^2}} }{\sqrt{1- {\varepsilon_{0-} (s)^2\over \Delta_0^2}} +{\zeta_0\over L}}\mbox{H}[\varepsilon_{0-} (s)]  \mbox{H}[\Delta_0 -\varepsilon_{0-}(s)]\right\} .~~~
           \eem{Jb}
Here  $\mbox{H}(q)$ are Heaviside step functions, which ensure that summation over $s$ extends only on states with energies $0<\varepsilon_{0\pm} <\Delta_0$ inside the gap. 

In the limit  $L\to 0$ (uniform superconductor without normal layer) the only contribution to the vacuum current is the current in the phase slip  state with the energy $\varepsilon_{ps}$ given by \eq{ASJ}: 
\be
J_v=J_{vA}=-{2e\over \hbar}{\partial  \varepsilon_{ps} (s)\over \partial \theta_0}={e\Delta_0 \over \hbar} \sin{\theta_0\over 2}.
     \ee{CT}

\subsection{Excitation current }\label{exc}
 
 The general expression for the excitation current in bound states is 
 \bem
J_{vA}=\sum_s \{2j _+(s)f_+(s)\mbox{H}[\varepsilon_{0+} (s)]  \mbox{H}[\Delta_0 -\varepsilon_{0+} (s)] 
\nonumber \\
+2j_-(s)f_-(s)\mbox{H}[\varepsilon_{0-} (s)]  \mbox{H}[\Delta_0 -\varepsilon_{0-} (s)]\},
    \eem{}
where $f_\pm(s)$ are occupation numbers of Andreev levels for rightmovers (+) and leftmovers (-). At nonzero temperature the occupation numbers are determined by the Fermi distribution 
\be
f_\pm(s)=\frac{1}{e^{\varepsilon_\pm(s)/T}+1}.
     \ee{}
At $T=0$ at the level with zero energy the occupation number  $f_\pm(s)$ is a free parameter in the interval from 0 to 1.

 As well as in  previous literature, at high temperatures (much higher that the Andreev level energy spacing 
 $ \pi \hbar v_f/L$, but still much lower than critical),  the present analysis focuses on the case of the long junction. Then one can ignore excitations in continuum states and replace sums for a large but finite number of Andreev states by infinite sums. 
 \bem
J_q=  \sum_{s=0}^\infty\frac{2j_+(s)}{e^{\varepsilon_+(s)/T}+1}+\sum_{s=0}^\infty\frac{2j_-(s)}{e^{\varepsilon_-(s)/T}+1}.
      \eem{JqS}

This expression is for a moving condensate, but one use $j_\pm(s)$ derived for the condensate at rest [\eq{js}]. This is because 
 a quasiparticle created in an Andreev state does not change the electron density \cite{SonAndr}.

Since  $\varepsilon_{0\pm}(s) \ll \Delta_0$ one may use \eq{enG}, and the excitation current is
\be
J_q={2ev_f\over L+\zeta_0} \sum\limits_0^\infty \left[\frac{1}{e^{\beta \left(s+{\pi+\theta \over 2\pi}\right)}+1}-\frac{1}{e^{\beta \left(s+{\pi-\theta \over 2\pi}\right)}+1}\right],
    \ee{JqIn} 
where
\be
\beta={\pi \hbar v_f\over (L+\zeta_0)T}.
  \ee{}
 While the vacuum current depends on the vacuum phase $\theta_0$, the excitation current is determined by  the total Josephson phase $\theta=\theta_0+\theta_s$.

\section{CPR at zero temperature} 

At zero temperature the CPR can be derived analytically without  calculations of sums and integrals for the vacuum currents  \cite{SonSh}.

At the Josephson phase $\theta$ less than critical one (see below) the total current reduces to the same condensate current in all layers as in a uniform superconductor. This conclusion is  valid beyond the steplike model for any thickness $L$ of the normal layer.

The condensate current is the only current flowing through the junction until the Landau criterion is satisfied.
The Landau criterion is violated when the energy of the lowest Andreev level $s=0$ for leftmovers (states with the excitation current flowing  in the direction opposite to the total current direction) vanishes. According to Eqs.~(\ref{GT}) and (\ref{eps0}), this happens when the phase $\theta=\theta_s$ reaches the critical value $\theta_{cr}$ determined by the equation
\be
\theta_{cr}={2L\varepsilon_{0-}(0,0)\over \hbar v_f} ={2m L\Delta_0 \over \hbar^2 k_f}\cos {\theta_{cr}\over 2}.
   \ee{thCr}
At $\theta >\theta_{cr}$ partial occupation of the lowest Andreev level starts and nonzero  phase $\theta_0$ appears. 

The CPR $J(\theta)$  at  $\theta >\theta_{cr}$ is derived from \eq{GT} at $\varepsilon_-=0$, \eq{eps00} at $s=0$,   and the relation \eq{Js} connecting the phase $\theta_s$ with the total current $J$: 
\be 
J= J_{cr}  \cos {\theta\over 2},
     \ee{}
where     
\be 
J_{cr}= {2 e\Delta_0 \over \pi \hbar} 
     \ee{Jcr}
is the critical current in the superconducting leads determined from the Landau criterion (depairing current).

Thus, there are two branches of the CPR.  At $\theta<\theta_{cr} $ the condensate current is the only current through the junction, and the phase distribution is the same as in a uniform superconductor. We call it the condensate current branch. Along this branch $\theta_0=0$ and $\theta=\theta_s$.  At  $\theta>\theta_{cr} $  the vacuum current and the excitation currents appear, but their sum vanishes, as required by the charge conservation law. At this branch $\theta_0\neq 0$ and $\theta_s\neq 0$. Along the branch the phase slip occurs when the phase difference across the junction loses  $2\pi$. So, the branch can be called phase slip branch.

\begin{figure}[!b] 
\centering
\includegraphics[width=0.35
 \textwidth]{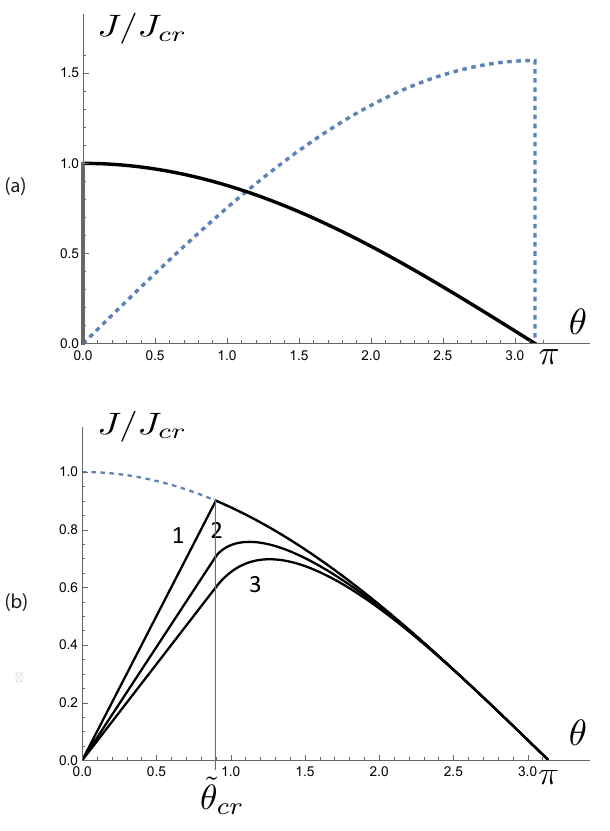} 
 \caption{CPRs at $T=0$. (a) $L=0$.   The  solid  line shows the CPR valid for any dimensionality of the junction. The current phase relation in the theory neglecting phase gradients in leads  is shown by the dashed line.
(b) $L=\tilde \zeta/2$. The curves 1, 2, and 3 are the current phase relations for  1D, 2D, and 3D junctions respectively. In the 1D case the length  $\tilde \zeta$ coincides with   $ \zeta_0$. \label{ShJ}}
 \end{figure}

The CPR at $L=0$, when the SNS becomes a uniform superconductor,  is shown in Fig.~\ref{ShJ}(a) by a solid line. Along the condensate current branch (vertical segment of the curve) the phase $\theta$ is equal to zero because there is no phase jump in a uniform superconductor in the subcritical regime. 
    
The CPR for a 1D junction with $L/\zeta_0=1/2$ is shown  in Fig.~\ref{ShJ}(b) (curve 1).  When the normal layer thickness $L$ grows the slope of the condensate current branch decreases and the branch approaches to the horizontal axis. The critical phase $\theta_{cr}$ approaches to $\pi$, while the phase slip branch becomes vertical. Eventually, the CPR is described by the saw-tooth current-phase curve (Fig.~\ref{ST}) for a very long junction. At growing $L$ the Josephson critical current (the maximal current across the junction) decreases from the bulk critical current $J_{cr}$ down to the very small current  $ev_f/L$. 

As was already mentioned, the derivation of the CPR does not require calculation of the vacuum current. This calculation is necessary only if one wants to know the occupation number in the phase slip state. We check occupation numbers required for the condition $J_v+J_q=0$ for the cases when there are  simple expressions for the  vacuum current $J_v$. In the limit $L=0$ the whole vacuum current is equal to the current  in the phase slip state given by \eq{CT}. The current $J_v+J_q$ vanishes if the occupation number is 1/2 along the whole  phase slip branch. In the  opposite limit $L\to \infty$ the total vacuum current is 
\be
J_v={ev_f\over L}{\theta_0\over \pi}.
     \ee{} 
On the other hand, the excitation current in the lowest Andreev level for leftmovers is (see Sec.~\ref{exc}) 
\be
J_q=2j_-(0)f=- {2ev_f\over L}f.
   \ee{}
Thus, $J_v+J_q=0$ if $f=\theta_0/2\pi$.
Since along the phase slip branch $\theta_0$ varies from 0 at $\theta=\theta_{cr}$ to $\pi$ at $\theta=\pi$, the occupation number varies from 0 to 1/2. At $\theta$ crossing $\pi$ the current changes its direction. Correspondingly, in this point the lowest Andreev level for leftmovers changes from half-occupied to empty, while  the lowest Andreev level for rightmovers changes from empty to half-occupied.
 
For short junctions the CPR taking into account phase gradients in leads is essentially different from that obtained ignoring these gradients \cite{ThunKink}.  The latter is shown by a dashed line in Fig.~\ref{ShJ}(a) at $L=0$.  This CPR is derived from the assumption that in shot junctions the total current reduces to the vacuum current  in the lowest Andreev level, determined by  \eq{CT} with $J_v=J$ and $\theta_0=\theta$ (no condensate current, $\theta_s=0$):
\be
J={e\Delta_0 \over \hbar} \sin{\theta\over 2}.
     \ee{KO}
This CPR was derived by Kulik and Omel’yanchouk \cite{KulOmel2} for non-planar junctions (narrow normal-metal bridges between  two bulk superconductors), which will be discussed in Sec.~\ref{bnd}.

 For multidimensional systems currents calculated for a single 1D channel  must be integrated over the space of wave vectors ${\bf k}_\perp$ transverse to the current direction keeping in mind that $k_f=\sqrt{k_F^2-k_\perp^2}$. Here $k_F$ is the radius of  the Fermi sphere in a multidimensional system. The integration operation is $\int_{-k_F}^{k_F}  {dk_\perp\over 2\pi}… $ in the 2D  case and $\int_0^{k_F}  {k_\perp\,dk_\perp\over 2\pi}… $ in the 3D case. After integration currents become current densities.

At the condensate current branch the linear CPR remains valid after integration, but one should replace the 1D density $n$ by the 2D or 3D density. The condensate current branch extends up the phase $\tilde \theta_{cr}$ determined by the equation
\be
\tilde\theta_{cr}- {2L \over\tilde \zeta}\cos {\tilde\theta_{cr}\over 2}=0,  
     \ee{}
which is similar  to \eq{thCr} for 1D junctions with $k_f$ replaced by $k_F$ and the 1D coherence length $\zeta_0$ [\eq{calL}] replaced by the   coherence length
\be
\tilde \zeta =  { \hbar^2 k_F \over  m \Delta_0} 
       \ee{}
 for multidimensional junctions. At $\theta =\tilde\theta_{cr}$ the transition to the phase slip branch occurs in the channel with the maximal $k_f=k_F$. In all other channels with $k_f< k_F$,  the condensate current branch extends up to phases larger than $\tilde \theta_{cr}$. Thus,  at  $\theta >\tilde \theta_{cr}$ we have a mixture of channels with the condensate current branch at $k_f<k_c$ and  with the phase slip branch at $k_f>k_c$. Here 
\be
k_c= {2L\cos {\theta\over 2} \over \tilde \zeta\theta }k_F.
 \ee{}
 Finally, integration over all channels yields 
\bem
J={2e\Delta_0 \over \pi^2 \hbar}\cos {\theta\over2}\int_0^{\sqrt{k_F^2-k_c^2}} dk+{e\hbar \over \pi^2 mL}\theta \int_{\sqrt{k_F^2-k_c^2}}^{k_F}\sqrt{k_F^2-k^2} dk
\nonumber \\
=J_{cr}\theta \left({\cos {\theta\over2}\over 2\theta}\sqrt{1-{4L^2\cos^2 {\theta\over 2} \over \tilde \zeta^2\theta^2 }}
+{\tilde \zeta \over 4L}\arctan{{2L\cos {\theta\over 2} \over \tilde \zeta \theta}\over \sqrt{1-{4L^2\cos^2 {\theta\over 2} \over \tilde \zeta^2 \theta^2}}}\right)
      \eem{2D}   
for the 2D junction and
\bem
J={e\Delta_0 \over \pi^2\hbar }\cos {\theta\over2} \int_0^{\sqrt{k_F^2-k_c^2}} k \,dk+{e\hbar \over 2\pi^2mL}\theta \int_{\sqrt{k_F^2-k_c^2}}^{k_F}\sqrt{k_F^2-k^2}2\pi k \,dk
\nonumber \\
=J_{cr}\cos {\theta\over2}\left(1-{4L^2\cos^2 {\theta\over 2} \over 3 \tilde \zeta^2\theta^2 }\right)
      \eem{3D}   
for the 3D junction. In the limit $L=0$ the expression for the ratio $J/J_{cr}$ in multidimensional junctions does not differ from that in 1D junctions, and the  plot $J/J_{cr}$ vs. $\theta$ [solid line in \ref{ShJ}(a)] describes the CPR for junctions of any dimensionality.  But  the critical current given by \eq{Jcr} for 1D junctions must be replaced by the critical current densities for multidimensional junctions:
\be
J_{cr}=\left\{ \begin{array}{cc}  {2 e\Delta_0 k_F \over \pi^2 \hbar}  &~~ 2D~\mbox{case} \\ \\  { e\Delta_0 k_F^2 \over 2\pi^2 \hbar} &~~  3D~\mbox{case} \end{array}\right. .
    \ee{}

The CPRs for 2D and 3D junctions at $L/\tilde \zeta=1/2$ are shown  in Fig.~\ref{ShJ}(b)  (curves 2 and 3) together with the CPR for a 1D junction (curve 1).  There is a cusp in the 1D CPR in the critical phase $\theta=\tilde\theta_{cr}$, which is smeared out in the 2D and 3D cases.  In 2D and 3D junctions the first derivative (slope) of the current-phase curve is continuous at $\theta=\tilde\theta_{cr}$, but the critical point is still non-analytic with jumps in a higher derivative.

\section{Long SNS junction  at high temperatures}

We consider high temperatures exceeding the Andreev level energy spacing $\pi \zeta_0/L$. Though the temperatures are called high, they are still much lower than critical ($T\ll \Delta_0$).
This temperature diapason is possible only in long junctions when $\pi \zeta_0/L \ll \Delta_0$. 

As discussed in Introduction, possible temperature independent corrections to the current decreasing faster than $1/L$ with growing $L$ are more important \cite{SonAndr} than the  exponentially decreasing with $T$ and $L$ current.  
But corrections $\propto 1/L^{3/2}$  and  $\propto 1/L^2$  calculated in Refs.~\cite{SonAndr,Son26}.vanished in the total current. For search of higher corrections numerical methods were used \cite{Son26}.

\begin{figure}[!t] 
\centering
\includegraphics[width=0.75 \textwidth]{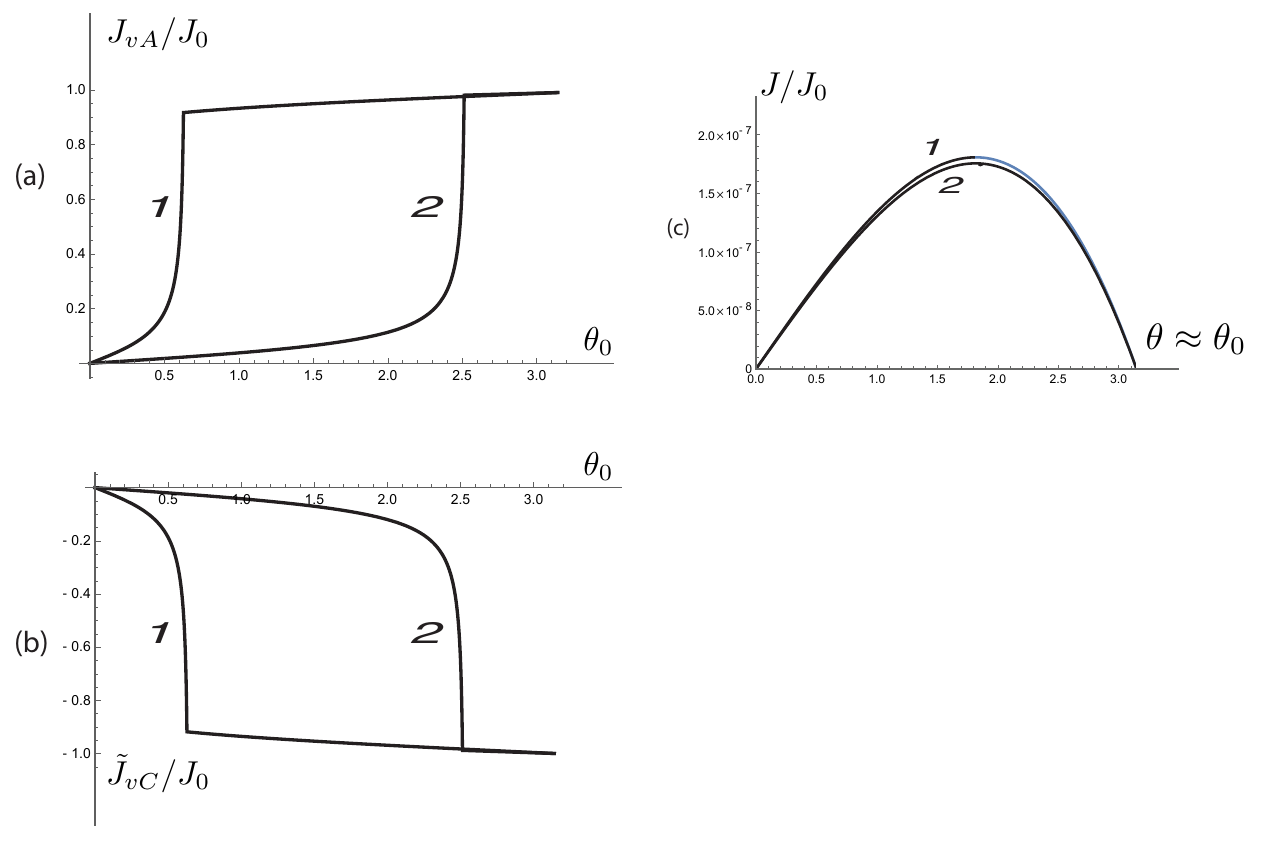} 
 \caption{The CPRs for 31 Andreev levels at $\alpha = 0.1$ (1) and $\alpha = 0.4$ (2).  (a) The vacuum current $J_{vA}$ in Andreev bound states. (b) 
 The reduced vacuum current $\tilde J_{vA}$  in continuum  states [\eq{JvCr}]. (c) The  total current $J$ at high temperatures.  
 \label{CPT}}
 \end{figure}

\subsection{Calculations of the vacuum current  in bound Andreev states}

At numerical calculation of the sum in \eq{Jb} for vacuum current  in bound Andreev states, energies of the states were determined by numerical solutions of \eq{eps00}.  Numerical calculations of the sum  with Mathematica encountered with  problems at large numbers of Andreev levels (large $L$), and some tricks were used. For example, the sum was divided on two or three sums calculated separately. The vacuum current in Andreev states $J_{vA}(\theta_0)$  is shown in Fig.~\ref{CPT}(a)  for $s_m=30$ [see \eq{inc}] for two values of $\alpha$. The cusp at $\theta_0=2 \pi\alpha$ is connected with the entrance or the exit of an Andreev level to or from the gap, which changes the number of Andreev levels from even to odd, or {\em vice versa} (parity effect \cite{SonAndr}). The cusp becomes a sharp current jump in the limit $L\to \infty$.

\subsection{Calculation of vacuum current  in continuum states}

For calculation of the continuum vacuum current we used the method described in Appendix B of Ref.~\cite{SonAndr}. In Ref.~\cite{SonAndr}  calculation was analytical  taking into account only the main terms $\propto 1/L$ and $\propto 1/L^{3/2}$. Now calculation is numerical using the exact {\em ab initio} expression [Eqs.~(\ref{Ta}) and (\ref{cJ})]. \footnote{In Ref.~\cite{Son21} the vacuum current  in continuum states was incorrectly ignored (see {\em Erratum} to Ref.~\cite{Son21}). The error was corrected in Ref.~\cite{SonAndr}. }

The continuum vacuum current is a difference  of contributions from right- and leftmovers:
\be
J_{vC} =J_+-J_- ,~~  J_\pm= - {e\over \pi\hbar}\int_{\Delta_0}^\infty  {\cal T}_\pm d\xi,  
     \ee{}
where ${\cal T}_\pm$ is given  by \eq{Ta}.

 Calculating $ J_\pm$ we introduce a new variable $z=(\varepsilon_0-\Delta_0)/\Delta_0$:
\be
J_\pm=  -{e\Delta_0\over \pi\hbar}\int_0^{x_m}\frac{2\sqrt{2z+z^2}(1+z)\,dz}{4z+2z^2+1-\cos(2Lz/\zeta_0+ \gamma_\pm)}, 
   \ee{jvc}
where
\be
\gamma_\pm=2\pi\alpha \mp \theta_0.
    \ee{}
The integrals for $J_\pm$ diverge, but their difference does not. So,  we introduced a  large $x_m$ as an upper cutoff assuming that in the end   $x_m\to \infty$. 

Integrands of the integrals are rapidly oscillating functions.  We  divide the whole interval of integration in \eq{jvc} on intervals of the length equal to the oscillation period  $\pi\zeta_0/L$ of the integrand:
\bem
J_\pm=-{e\Delta_0\over \pi \hbar}\left[I_0(\gamma_\pm)+ \sum_{s=1}^{s_{xm}} I(s,\gamma_\pm)+I_m(\gamma_\pm) \right],
     \eem{jcp}
where
\be
I(s,\gamma_\pm)
=\int_{z_{s\pm}-{\pi\zeta_0\over 2L}}^{z_{s\pm}+{\pi\zeta_0\over 2L}} \frac{2\sqrt{2z+z^2}(1+z)\,dz}{4z+2z^2+1-\cos(2Lz/\zeta_0+ \gamma_\pm)}.
   \ee{is1}
Here
\be
z_{s\pm} ={\zeta_0\over 2L}(2\pi s -\gamma_\pm)
    \ee{poles}
 is the coordinate of the period center, where the cosine argument is $2\pi s$, 
 and $s_{xm}$ is the maximal integer $s$ satisfying the condition that the both upper limits 
 $z_{s\pm}+{\pi\zeta_0\over 2L}$ are smaller than $x_m$. The integrand in \eq{is1} is equal to 1 at  very large $z$. At large $L$ and small $z$  the integrand becomes a sharp peak at $z=z_{s\pm}$. These peaks are  peaks of the transmission  probability (transmission resonances) \cite{Bard,Svidz71,Kummel}. The energy spacing between transmission resonances is the same as that for bound Andreev levels.
 
 The term  $I_m(\gamma_\pm)$ takes into account the integration interval between the upper limit $x_m$ in the integral \eq{jvc} and  the upper border of the last $s=s_{xm}$  period. Since the integrand at large $z$ goes to 1,  the term can be calculated exactly:
\be
I_m(\gamma_\pm)= x_m-  {(2\pi s_{xm}+\pi-\gamma_\pm)\zeta_0\over 2L}. 
     \ee{Im} 
 
The term  $I_0(\gamma_\pm)$ is the contribution of the integration interval between $z=0$ and  the lower  border of the $s=1$ interval. The value of  $I_0(\gamma_\pm)$  is determined by the integral in \eq{is1} at $s=0$ with the lower limit replaced by 0.

Finally,  the  continuum vacuum current is
\bem
J_{vC} =\tilde J_{vC} -{e\Delta_0\over \pi \hbar}\left[I_m(\gamma_+)-I_m(\gamma_-) \right]=\tilde J_{vC} +  {ev_f\over L}{\theta_0 \over \pi},
      \eem{Jvc}
where only the reduced continuum vacuum current 
\bem
\tilde J_{vC} =-{e\Delta_0\over \pi \hbar}
\left\{I_0(\gamma_+)-I_0(\gamma_-)+ \sum_{s=1}^{s_{xm}} [I(s,\gamma_+) -I(s,\gamma_-)]\right\}
     \eem{JvCr}
must be calculated numerically. Below we shall see that the largest term  $\propto 1/L$ in \eq{Jvc} is compensated in the total current by the similar term in the excitation current with the opposite sign.

The reduced vacuum current in continuum states $\tilde J_{vC}(\theta_0)$  is shown in Fig.~\ref{CPT}(b)  for $s_m=30$ for two values of $\alpha$. The cusp at $\theta_0=2 \pi\alpha$ has an opposite sign to the cusp in bound states, and there is no cusp in the total current (see Sec.~\ref{curPh}).

\subsection{Calculation of excitation current}

For temperatures much larger  than the energy spacing between Andreev levels but much smaller than the gap, the sum in the expression \eq{JqIn} for the excitation current can be calculated analytically \cite{Bard,SonAndr}:
\bem
J_q=-{ev_f\over L+\zeta_0}{\theta \over \pi}+{8e T\over \hbar } e^{-2\pi TL/\hbar v_f}\sin \theta.
     \eem{JqM}
The first term is the expression \eq{JqIn} with the sum replaced by integration over $s$. The second exponential term is the contribution of Matsubara poles emerging at the exact calculation of the sum.

Since we subtracted the term $\propto 1/L$ from the vacuum current, we should also subtract   the same term with opposite sign in the excitation current. Thus, we introduce the reduced excitation current   
\be
\tilde J_q=J_q+{ev_f\over L}{\theta \over \pi}={ev_f \over L}{\zeta_0\over L+\zeta_0}{\theta \over \pi}.
    \ee{q2}
The exponentially small  Matsubara term  in \eq{JqM} was neglected.

The final value $\propto 1/L^4$ of the current is very small, and  subtraction of larger terms $ \propto 1/L$ from sums makes the problem of small difference between large terms easier.  But this does not eliminate it completely. In calculated sums terms much larger than the total  current still  remain. 

\subsection{CPR}\label{curPh}

At high temperatures the SNS junction is a weak link, and the superfluid phase $\theta_s$ is very small ($\theta \approx \theta_0$). Then phase gradients in leads can be ignored, as was done in the past. The total current is determined by currents flowing in the normal layer:
\be
J=J_{vA}+\tilde J_{vC}+ \tilde J_q.
     \ee{Jt}

The numerically calculated current $J$ is shown in Fig.~\ref{CPT}(c)  for $s_m=30$.  The total current $J$ is by the factor about $\sim 10^{-7}$ smaller than the currents in the sum determining $J$. This illustrates the problem of small difference between large terms mentioned above. 

\begin{figure}[!t] 
\centering
\includegraphics[width=0.4 \textwidth]{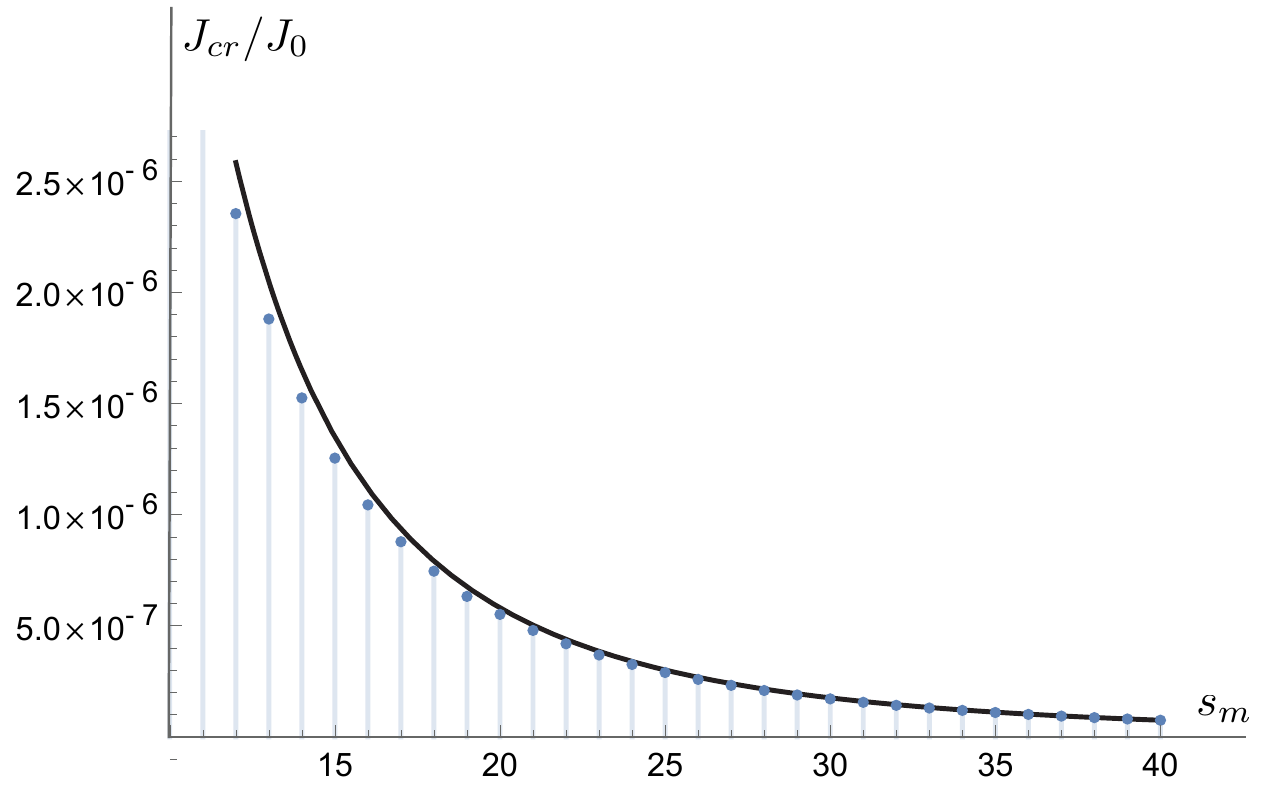} 
 \caption{Josephson critical current vs. number of Andreev levels $s_m$ for $\alpha=0.4$ (discrete plot). The continuous solid line shows the power law $0.005/s_m^3$.    \label{PL}}
 \end{figure}

According to Fig.~\ref{CPT}(c), the dependence on $\alpha$ is rather weak. It  is difficult to decide whether it   really exists but numerically small,  or is connected with   numerical inaccuracy. Anyway, this dependence can be ignored.  The CPR is close (but not identical) to sinusoidal CPR 
\be
J=J_J \sin\theta.
   \ee{HT}
The Josephson critical current $J_J$  vs. the number $s_m$  is shown in Fig.~\ref{PL}. The plot is close to $J_J=0.005 J_0/s_m^3$. According to Eqs.~(\ref{JL})
and (\ref{inc}) for $J_0$ and $s_m$, the Josephson critical current is
\be
J_J= 0.15 {ev_f\over L}{\zeta_0^3\over L^3}=0.15{e\hbar^3\over \Delta_0^3}   {v_f^4\over L^4}.
      \ee{cr}

At large $L$ and high temperatures the temperature independent  power law current exceeds the exponentially decreasing Matsubara term in \eq{JqM}. Thus, the prediction of exponentially small critical current at high temperature in some publications  starting from Bardeen and Johnson \cite{Bard} must be revised for planar junctions. The exponential law is  still valid for not very high temperatures less than $T^*$ \cite{SonAndr}. Comparing the power law current in \eq{HT} with 
the exponentially small term  in \eq{JqM} the temperature $T^*$ is
\be
T^*={3\hbar v_f\over 2\pi L}\ln{L\Delta_0\over \hbar v_f}.
     \ee{}
In this expression we ignored numerical factors in the logarithm argument as not important in the limit $L\to \infty$. 

\part{Beyond planar SNS junctions with equal Fermi velocities} \label{part2}

If Fermi velocities  in superconductors and a normal metal are not equal, the normal scattering cannot be ignored in general.  But in the limit $L \to 0$ when the normal layer disappears,   the CPR cannot depend on the Fermi velocity and the effective  mass of the absent normal metal. This means that the CPR at $L=0$ shown in Fig.~\ref{ShJ}(a) by solid line is valid in this case. Further we focus on short junctions with small but finite $L$.

\section{NNN  junctions with non-equal Fermi velocities}

In order to understand what is going on in SNS junctions with non-equal Fermi velocities at finite $L$ it is useful to consider a much simpler problem of the NNN junction  when superconducting leads are replaced by normal metals.

A 1D normal wire has a barrier (or well) produced  by an external rectangular potential between $x=-L/2$ and $x=L/2$. The Fermi levels are above  the barrier everywhere, so there is the Fermi sea in all wire segments with the Fermi wave numbers $k_f$ outside the barrier and $k_0=\sqrt{k_f^2-2me\Phi/\hbar^2}$  
inside the barrier. The electrostatic scalar potential $\Phi$ can be created by a gate.
The barrier height $e\Phi$ is negative in the case of well.  Further we focus on the case of barrier ($k_f>k_0$). 

The wave function for a particle incident from the left ($x=-\infty$)  taking into account reflection is  
\be
\Psi =\left\{ \begin{array}{cc} e^{ ik_f x}+r_n e^{ -ik_fx} & x<-L/2 \\  Ae^{ ik_0 x}+ A'e^{ -ik_0x} & -L/2<x<L/2\\ t_ne^{ ik_f x} & x>L/2
      \end{array} \right. .
   \ee{}
Here $r_n$ and $t_n$ are reflection and transmission amplitudes at normal  scattering. They, together with the parameters $A$ and $A'$ of the wave function in the barrier area, are determined by matching the wave function and its derivative at the interfaces $x=\pm L/2$. This yields values of reflection and transmission probabilities:
\be
{\cal R}=|r_n|^2=\frac{2[1-\cos (2k_0L)]}{G^2+2[1-\cos (2k_0L)]},~~
{\cal T}=|t_n|^2=\frac{G^2}{G^2+2[1-\cos (2k_0L)]}.
      \ee{}
Here
\be
G={4k_fk_0\over k_0^2-k_f^2}
  \ee{}
is the parameter characterizing the mismatch  between Fermi velocities $\hbar k_f/m$ and  $\hbar k_0/m$. Note the symmetry between the cases $k_0 < k_f$ (barrier) and  $k_0 > k_f$ (well). It is remarkable  that the junction becomes fully transparent (${\cal T}=1$) at discrete  values $k_0=\pi s/L$ ($s$ is an integer), which correspond to Fabry–P\'erot resonances.

Below the Fermi levels the current in the wave propagating to the right  (rightmover) is compensated by the current  in the wave propagating to the left (leftmover). But applying the voltage $V$ to leads one raises the Fermi level for rightmovers ($k_f \to k_f +eVm/2\hbar^2k_f$) and lowers the Fermi level for leftmovers  ($k_f \to k_f -eVm/2\hbar^2k_f$). There are only rightmovers in the interval of wave numbers between  $ k_f -eVm/2\hbar^2k_f$ and $ k_f +eVm/2\hbar^2k_f$ They produce the total current. At small voltage one can neglect  the difference of wave numbers from the Fermi wave number, and the total current  is (taking into account spin):

\be
J={e^2V\over \pi \hbar}{\cal T}={e^2V\over \pi \hbar}\frac{G^2}{G^2+2[1-\cos (2k_0L)]}.
    \ee{}
This means that at small $G$ (strong mismatch of Fermi velocities) the resistance of the NNN junction is
\be
R_J ={2\pi \hbar\over e^2G^2}[1-\cos (2k_0L)] ={\pi \hbar\over 8 e^2}{k_f^2\over k_0^2}[1-\cos (2k_0L)].
       \ee{RJ}

\section{Planar SNS junctions with normal scattering (non-equal Fermi velocities in leads and inside the junction)}

\subsection{Delocalized continuum states}

Now we return back to the SNS junction. The  expression for the  Bogolyubov\,\textendash\,de Gennes wave function of a rightmover quasiparticle   taking into account normal scattering are [instead of Eqs.~(\ref{BGr}--\ref{norm})] 
\bem
\left(  \begin{array}{c}
 u    \\
  v 
\end{array}  \right)=\left(  \begin{array}{c}
U_+  e^{-i\theta_0/4+i \nabla \varphi\,x/2}  \\
 V_+  e^{i\theta_0/4-i \nabla \varphi\,x/2}
\end{array}  \right)e^{ i\left(k_f+{ m\xi_+\over \hbar^2k_f}\right)x}
\nonumber \\ 
 +r_c \left(  \begin{array}{c}
U_-e^{-i\theta_0/4+i \nabla \varphi\,x/2}   \\
 V_- e^{i\theta_0/4-i \nabla \varphi\,x/2} 
\end{array}  \right)e^{-i\left(
k_f+{ m\xi_-\over \hbar^2k_f}\right)x}
 +r_a \left(  \begin{array}{c}
V_+e^{-i\theta_0/4+i \nabla \varphi\,x/2}   \\
U_+ e^{i\theta_0/4-i \nabla \varphi\,x/2} 
\end{array}  \right)e^{i\left(k_f-{ m\xi_+\over \hbar^2k_f}\right)x}
      \eem{Gr}
for $x<-L/2$, 
\bem
\left(  \begin{array}{c}
 u    \\
 v 
\end{array}  \right)=t_c\left(  \begin{array}{c}
U_+ e^{i\theta_0/4+i \nabla \varphi\,x/2}  \\
 V_+e^{-i\theta_0/4-i \nabla \varphi\,x/2}
\end{array}  \right)
e^{ i\left(k_f+{ m\xi_+\over \hbar^2k_f}\right)x}
+t_a\left(  \begin{array}{c}
V_- e^{i\theta_0/4+i \nabla \varphi\,x/2}  \\
 U_-e^{-i\theta_0/4-i \nabla \varphi\,x/2}
\end{array}  \right)
e^{- i\left(k_f-{ m\xi_-\over \hbar^2k_f}\right)x}
      \eem{Gt}
for $x>L/2$,
and
\be
  \left(  \begin{array}{c}
 u    \\
 v 
\end{array}  \right)= \left( \begin{array}{c} Ae^{ik_0x+im\varepsilon x/\hbar^2k_0}+A'e^{- ik_0x-im\varepsilon x/\hbar^2k_0}  \\  
 Be^{ik_0x-im\varepsilon x/\hbar^2k_0}
 +B'e^{- ik_0x+im\varepsilon x/\hbar^2k_0}
 \end{array}  \right)
              \ee{NA}
for the normal area $-L/2<x<L/2$. 
Here 
\be
\varepsilon_\pm= \varepsilon \mp {\hbar^2 k_f \over 2m}\nabla \varphi,~~\xi_\pm=\sqrt{\Delta_0^2-\varepsilon_\pm^2},
      \ee{}
are Doppler-shifted energies for rightmovers and leftmovers, and we introduced a smaller number of wave functions parameters [instead of those in \eq{u0v0}]:
\bem
U_\pm=\sqrt{\varepsilon_\pm +\xi_\pm\over 2\varepsilon_\pm},~~V_\pm=\sqrt{\varepsilon_\pm -\xi_\pm\over 2\varepsilon_\pm}.
    \eem{UV}

In the presence of normal scattering not only all components of wave functions but also their derivative must be continuous at the interfaces $x=\pm L/2$. One can find the scattering parameter $A$, $A'$, $B$, and $B' $ in the normal area as linear functions of $t_c$ 
and $t_a$ from the boundary conditions at $x=L/2$ and as linear functions of $r_c$ 
and $r_a$ from the boundary conditions at $x=-L/2$. From the condition that $A$, $A'$, $B$, and $B' $ found by two methods must coincide one obtains four linear equations for scattering parameters $r_c$, $r_a$, $t_c$, and $t_a$:

 \bem
(k_0+k_f)U_++(k_0-k_f)U_-\tilde r_c+(k_0+k_f)V_+\tilde r_a
\nonumber \\
= e^{i(\mp k_0L+\Theta)/2} ((k_0+k_f) U_+\tilde t_c +(k_0-k_f)V_-\tilde t_a)
\nonumber \\
(k_0-k_f)U_++(k_0+k_f)U_-\tilde r_c+(k_0-k_f)V_+\tilde r_a 
\nonumber \\
=e^{i(\pm k_0L+\Theta)/2} ((k_0-k_f)U_+\tilde t_c +(k_0+k_f)V_-\tilde t_a)
\nonumber \\
(k_0+k_f)V_++ (k_0-k_f)V_-\tilde r_c + (k_0+k_f)U_+\tilde r_a    
\nonumber \\
=e^{i(\mp k_0L-\Theta)/2}((k_0+k_f) V_+\tilde t_c +(k_0-k_f) U_-\tilde t_a)
 \nonumber \\
(k_0-k_f)V_++ (k_0+k_f)V_-\tilde r_c + (k_0-k_f)U_+\tilde r_a  
\nonumber \\
=e^{i(\pm k_0L-\Theta)//2}((k_0-k_f)V_+\tilde t_c +(k_0+k_f)U_-\tilde t_a)
      \eem{ab}
where
\be
\tilde r_c=r_ce^{ik_fL},~~\tilde r_a=r_a,~~\tilde t_c=t_c e^{ik_fL}
,~~\tilde t_a=t_a.     \ee{}
Phase factors of unit modulus connecting tilded and non-tilded scattering parameters do not affect reflection and transmission probabilities ${\cal R}_c=|\tilde r_c|^2$, ${\cal R}_a=|\tilde r_a|^2$, ${\cal T}_c=|\tilde t_c|^2$, and ${\cal T}_a=|\tilde t_a|^2$.

In these equations we ignored $v_s$- and $\xi_\pm$-dependent terms to the wave numbers. These terms decrease as $1/k_f$ with growing $k_f$ and are not important in the weak coupling  limit $\Delta \ll \varepsilon_f$. Still, we kept terms $\sim \xi/k_f$ in the exponent arguments  because they determine the group velocity of quasiparticles and quasiholes and demonstrate the difference between various wave components in the wave. superposition for scattering states. The main effect of the condensate motion is determined by the Doppler  shifts of energies. Since we are going to use these equations for short junctions with $L$  much smaller than the coherence length $\zeta_0$, we replaced $k_0\pm m\varepsilon /\hbar^2k_0$ by $k_0$ in arguments of exponential functions.

A junction without normal scattering becomes a uniform superconductor at $L$ smaller than the coherence length $\zeta_0$.  But we shall  see that at strong mismatch of Fermi velocities transformation to  a uniform superconductor takes place at much shorter scale.
At $L \ll \zeta_0$,  the solution of 4 linear equations in \eq{ab} is [after replacing the wave function parameters $U_\pm$ and $V_\pm$ by their expressions in \eq{UV}]:
\bem
r_c= -\frac{\sqrt{(\varepsilon_++\xi_+)(\varepsilon_-+\xi_-)} -\sqrt{(\varepsilon_+-\xi_+)(\varepsilon_--\xi_-) }     }{2\sqrt{\varepsilon_+\varepsilon_-}D}\{\sqrt{G^2+4}[\cos (2k_0L)-1] +Gi\sin k_+ \},
     \eem{rc}\bem
r_a=  -{\Delta_0\over 2\varepsilon_+D}G^2[[(\cos\theta-1)-{\xi_-\over \varepsilon_-}i\sin \theta] +2\left({\Delta_0\over 2\varepsilon_-} - {\Delta_0\over 2\varepsilon_+}\right)[\cos (2k_0L)-1],
  \eem{ra}
\bem
t_c    ={\xi_+\over \varepsilon_+D}G^2 \{- {\xi_-\over \varepsilon_-}\cos (k_0L)\cos(\theta/2) 
  + \cos (k_0L)i\sin(\theta/2)\} 
   \nonumber \\%
+{\xi_+\over \varepsilon_+}G\sqrt{G^2+4} \{ -{\xi_-\over \varepsilon_-}i\sin(k_0L)\cos(\theta/2) 
  - \sin(k_0L)\sin(\theta/2)\}, 
  \eem{tc}
\bem
t_a= -2iG{\xi_+\over \varepsilon_+D} \left[\frac{\sqrt{(\varepsilon_++\xi_+)(\varepsilon_--\xi_-)} -\sqrt{(\varepsilon_+-\xi_+)(\varepsilon_-+\xi_-) }     }{2\sqrt{\varepsilon_+\varepsilon_-}}\cos(\theta/2) 
\right.  \nonumber \\ \left.
+\frac{\sqrt{(\varepsilon_++\xi_+)(\varepsilon_--\xi_-)} +\sqrt{(\varepsilon_+-\xi_+)(\varepsilon_-+\xi_-) }     }{2\sqrt{\varepsilon_+\varepsilon_-}}i\sin  (\theta/2)\right]\sin(k_0L),
   \eem{ta}
\bem
D=G^2 [ {\varepsilon_+\varepsilon_- -\xi_+\xi_-\over 2\varepsilon_+\varepsilon_-}\cos\theta+{\xi_+\varepsilon_- -\varepsilon_+\xi_-\over 2\varepsilon_+\varepsilon_-}i\sin\theta]
 +2 {\varepsilon_+\varepsilon_- +\xi_+\xi_--\Delta_0^2\over 2\varepsilon_+\varepsilon_-} [\cos (2k_0L)-1]
    \nonumber \\
 -  G^2 {\varepsilon_+\varepsilon_- +\xi_+\xi_-\over 2\varepsilon_+\varepsilon_-} .
   \eem{de}
One can check that at these scattering parameters the quasiparticle flux \eq{g} remains constant along the whole junction.

Similarly, one obtains scattering parameters for other types of scattering states. Parameters for a rightmover quasihole incident from right are given by  Eqs.~(\ref{ra}--\ref{de}) after changing signs before $\xi_\pm$, $k_0$, and $\theta$. For a leftmover quasiparticle one should change signs of $\theta$ and subscript $\pm$ ($\pm\to \mp$). Correspondingly, for a leftmover quasihole one should change signs before $\xi_\pm$ and $k_0$, and replace $\pm$ by $\mp$.

\begin{figure}[!t]
\centering
\includegraphics[width=.7\textwidth]{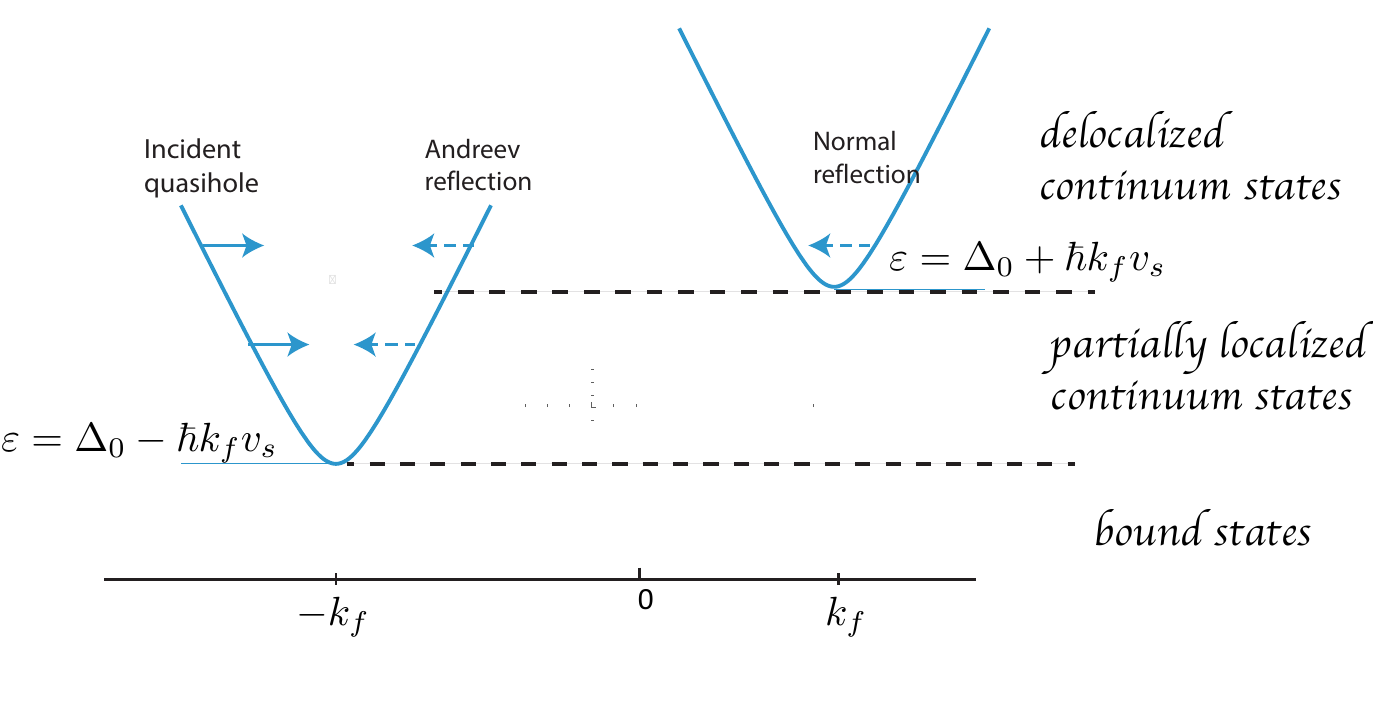}
\caption[]{Scattering of the leftmover quasihole. Its normal scattering is impossible at energies between Doppler-shifted gaps.  Note that arrows show directions  of quasiparticle and  quasihole motion, i.e., of group velocity, but mot of wave vector. So, the incident leftmover quasihole moves to the right.      \label{PLS} }
\end{figure}

\subsection{Partially localized continuum states }

Above we consider scattering states, which were superpositions only of propagated plane waves. But at non-equal Fermi velocities there is an energy interval, where the superposition should include not only propagated waves but also bound state  components, which penetrate into leads as evanescent waves. We call these scattering states partially localized continuum states. Despite presence of a localized component (bound state component), these states have continuous energy spectrum, i.e., they belong to the space of continuum  states.

Inevitability of partially localized continuum states is evident from simple energy arguments illustrated in Fig.~\ref{PLS}. In the presence of phase gradients (moving condensate) the Doppler shifts have opposite signs for rightmovers and leftovers. In the energy interval between two Doppler-shifted gaps there is no incident rightmover quasiparticle or quasihole . At the same time, normal scattering of  leftmover quasiparticle or quasihole becomes impossible since the propagating normal-scattered wave becomes an evanescent wave in superconducted leads. Thus, this wave component is localized in the normal layer. At transition from delocalized continuum states
to partially localized,  the real energy $\xi_+=\sqrt{(\varepsilon_+-\hbar k_fv_s)^2-\Delta_0^2}$ becomes imaginary:
$\xi_+=i\sqrt{\Delta_0^2-(\varepsilon_+-\hbar k_fv_s)^2}$.

\subsection{Bound states} \label{BSC}

Energies of bound states in the normal layer are zeros of the determinant of equations in \eq{ab}. Further our analysis will focus on the case of strong mismatch of Fermi velocities when the SNS junction becomes a weak link, and phase gradients in leads are small, i.e.,  
$\varepsilon_+\approx \varepsilon_-$, $\xi_+\approx \xi_-$. The equation for bound state energy is
 \bem
 D\propto G^2  (\Delta_0^2\cos\theta -2\varepsilon^2+\Delta_0^2) - 4(\varepsilon^2-\Delta_0^2) [1-\cos (2k_0L)]=0
       \eem{}
The energy of the bound state is
\be
\varepsilon=\Delta_0\sqrt{ G^2  \cos^2{\theta\over 2}  +2[1-\cos (2k_0L)] \over  
G^2  + 2[1-\cos (2k_0L)]}.
   \ee{BSe}
As expected, at $L=0$ this yields the energy of the bound state obtained for short junctions without normal scattering [\eq{ASJ}]. However, at strong mismatch of Fermi velocities (small $G$) corrections to this energy at finite $L$ start not at $L$ of the order of the coherence $\zeta_0$,  but at much shorter scale on the order of interparticle distance $1/k_f$ in leads. This follows from comparison of two terms in the denominator in \eq{BSe}. 

In the limit of strong mismatch  when the term $\propto G^2$ is smaller than the $L$-dependent term  
\be
\varepsilon=\Delta_0\left\{1 -{G^2  (1-\cos\theta) \over 8[1-\cos (2k_0L)]}\right\}.
         \ee{Ben}

\subsection{CPR at strong mismatch of Fermi velocities}

The vacuum current in continuum states is determined by the expression similar to that in \eq{cJ}. Now it should include transmission  
probabilities connected both with Andreev and normal scattering. At  strong mismatch of Fermi velocities the SNS junction becomes a weak link, and the previous approach ignoring phase gradients in leads at calculation of  the current in the normal layer becomes legitimate. So, we may simplify Eqs.~(\ref{rc}--\ref{de}) ignoring Doppler shifts. For calculation of the vacuum current in continuum states in the normal layer it is sufficient to know only transmission amplitudes:
\bem
t_c    ={4\xi\over \varepsilon D}G \left(-{\xi\over \varepsilon}i\sin(k_0L)\cos{\theta \over 2}  - \sin(k_0L)\sin{\theta \over 2} \right) ,
 \nonumber \\
 t_a= \frac{\Delta_0 }{\varepsilon D}i\sin {\theta \over 2} \sin(k_0L),
 \nonumber \\
 D=G^2  \left({\Delta_0^2\over 2\varepsilon^2}\cos\theta-  {\Delta_0^2+2\xi^2\over 2\varepsilon^2} \right)+2 {\xi^2\over \varepsilon^2} [\cos (2k_0L)-1].
    \eem{de0}
According to these expressions, transmission probabilities are even functions of $\theta$. Meanwhile, the current is an odd function of $\theta$. Thus, the continuum vacuum  current   determined by even in $\theta$ transmission probabilities must vanish.

The only current in the junction is the current  in the bound state discussed in Sec.~\ref{BSC}. Using the canonical relation connecting 
the current with the derivative of the energy in \eq{Ben} with respect to the phase, the total current is
\be
J=-{2e\over \hbar}{\partial  \varepsilon \over \partial \theta}={e\Delta_0 \over 4\hbar} {G^2  \sin\theta \over 1-\cos (2k_0L)}.
     \ee{}
Comparing this with the resistance of the NNN junction given by \eq{RJ} we obtain the Ambegaokar–Baratoff relation:
\be
J_{max}R_J={\pi \Delta_0 \over 2e},
     \ee{}
where $J_{max}$ is the maximal current  in the junction at $\theta=\pi/2$. Thus, at strong mismatch of Fermi velocities the short ballistic SNS junction becomes a common weak-link Josephson junction.

\section{Beyond planar junctions (narrow bridges between leads of higher dimensions)} \label{bnd}

Let us discuss possible implications of the present analysis  for a non-planar junction: a narrow normal-metal bridge between bulk superconducting leads [Fig.~\ref{pg}(b)]. It was considered  by Kulik and Omel’yanchouk \cite{KulOmel2}  using the Green function formalism. In contrast to a planar junction, in the limit  $L\to 0$ the bridge junction becomes not a uniform superconductor, but a superconductor divided by a thin impermeable  partition wall with a small orifice. Despite this difference, the theory ignoring phase gradients predicts the same CPR given by \eq{KO}.
Let us explain this  within our approach.

The current in \eq{KO} is the vacuum current in the phase slip bound state described by two evanescent waves on both sides from the phase slip point (small orifice in the case of  Kulik and Omel’yanchouk). The evanescent waves are plane waves in a planar junction, but are cylindrical waves in the 2D case or spherical waves  in the 3D case. 
Kulik and Omel’yanchouk considered orifices of diameter much large than the interparticle distance $\sim 1/k_f$ (but less than the coherence length). In this case cylindrical and spherical  waves are described by asymptotic expressions for zero orbital moment (no angular dependence): $e^{(ik_f -1/\zeta)r}/\sqrt{r}$ in the 2D case and $e^{(ik_f -1/\zeta)r}/r$  in the 3D case. Here $r$ is the distance from the small orifice. Divergence at $r\to 0$ can be cut off by the size of the orifice. Replacing plane waves by cylindrical or spherical  waves does not affect the condition of the continuity of two evanescent waves  [\eq{CE}]. So, our approach presents a clear physical picture of the vacuum current through a short non-planar junction, which is not evident in the sophisticates Green function analysis \cite{KulOmel2,Green}. 

Kulik and Omel’yanchouk \cite{KulOmel2} ignored phase gradients in leads as was common in the past. Far from the orifice phase gradients are very small indeed. But at approaching the orifice they grow as $1/r$ in the 2D case and as $1/r^2$ in the 3D case. Close to the orifice the current in the bridge requires the same gradients as in a planar junction.  The analysis of Kulik and Omel’yanchouk \cite{KulOmel2} widely accepted up to now (see, e.g., Eq.~(7) in Ref.~\cite{ThunKink}, Eq.~(177) in Ref.~\cite{Green}, or Eq.~(3.3) in Ref.~\cite{JJbk}) rules out the existence of the condensate current branch with zero phase jump  at the orifice.

Kulik and Omel’yanchouk assumed  that the gap  $\Delta_0$ is constant everywhere ignoring its possible suppression near the orifice. At this assumption one may expect a current state of an ideal incompressible fluid flowing through the orifice, which is not accompanied by phase jump at the orifice. Thus, even in the orifice case one may expect in the limit   $L\to 0$  the backward-skewed  CPR [solid line    in Fig.~\ref{ShJ}(a)] rather than the forward-skewed CPR obtained ignoring the phase gradients in leads [dashed line    in Fig.~\ref{ShJ}(a)]. 

Observation of the backward-skewed CPR in short bridge non-planar SNS junctions \cite{InAs} may be considered as  evidence in favor of this scenario, which connects  backward-skewness in Ref.~\cite{InAs}  with the effect of phase gradients in leads \cite{Krekels}.
But further quantitative analysis of non-planar junctions  taking into account, in particular, normal scattering is needed for final conclusion about the nature of  the backward-skewed  CPR in the experiment \cite{InAs}. This  is beyond the scope of the present work.

One aspect of the experiment on short bridge non-planar SNS junctions  by Spanton {\em et al.} \cite{InAs} requires further discussion.
They stressed that they observed a backward-skewed CPR only for a narrow gate voltage range in the gated junction, and never in ungated junctions. 
This looks as a contradiction to our analysis of planar junctions predicting the backward-skewed CPR only in ungated junctions. In fact, there is no contradiction. 

In ungated bridge junctions there is no electrostatic potential creating a barrier located in the normal area. But there is another energy, which raises the bottom of the Fermi sea in the normal area (bridge) with respect to the bottom of the Fermi sea in the leads. This is the energy necessary for confinement of electrons in a narrow bridge: $E_{cf} \sim \hbar^2/md^2$, where $d$ is on the order of thickness of the  bridge. Adding a gate, one can reach the situation when the electrostatic potential created by the gate fully compensates the effect of the confinement energy: $e\Phi+E_{cf}=0$. Then the backward-skewed CPR emerges.

These qualitative arguments cannot replace a quantitative analysis of non-planar junctions taking into account phase gradients in superconducting leads.  This analysis is expected to be more complicated than that for planar junctions. It is beyond the scope of the present article.

\section{Conclusions}

The recently suggested  theory of ballistic SNS junctions reviewed in this article  demonstrates that  planar junctions with equal 
Fermi velocities in all layers at $T=0$ are not weak links, and phase gradients in leads cannot be ignored, as commonly done in the past. At $T=0$ the  CPR was derived analytically  for any normal layer thickness $L$ in the steplike pairing potential model. This derivation relied on the Galilean invariance in the SNS junction without normal scattering and the Landau criterion for bound Andreev states. The suggested theory resolves the problem with the charge conservation law within the steplike pairing potential model commonly used during a half century. The theory predicts the backward-skewed CPR in short junctions, which was not expected in the previous theory. The backward-skewed CPR was reveals in numerical \cite{Krekels} and physical \cite{InAs} experiments.

The suggested  theory revised also the previous theory for the long ballistic SNS junction for temperatures much higher than  the Andreev level energy spacing (but still much lower than critical). Instead of exponential decrease of the maximal  current through the SNS junction with growing $L$ and $T$ in the previous theory, our analysis predicts the temperature independent plateau and  power law $1/ L^4$ decrease  with growing $L$. 

Planar junctions with non-equal Fermi velocities in superconductors and a normal layer were also analyzed.  In this case normal scattering cannot be ignored. At strong mismatch of Fermi velocities the planar ballistic SNS junction becomes a common Josephson
junction, which is a weak link and satisfies the Ambegaokar–Baratoff relation connecting the
maximal current through the junction with the resistance of the corresponding normal (NNN)
ballistic junction.

Although the quantitive analysis was done for planar SNS junctions, the analysis has implications  for non-planar junctions [narrow normal bridges between bulk superconducting leads, Fig.~\ref{pg}(b)]. The article  suggests an explanation why the    backward-skewed CPR  in short InAs nanowire bridge junctions \cite{InAs}  was observed only for
a narrow gate voltage range in the gated junction.


\providecommand{\newblock}{}

\end{document}